\documentclass[10pt, conference, compsocconf]{IEEEtran}

\usepackage{cite}

\usepackage[english]{babel}
\usepackage[utf8x]{inputenc}
\usepackage[T1]{fontenc}

\usepackage{amsmath}
\usepackage{graphicx}

\usepackage{algorithm}
\usepackage[noend]{algpseudocode}

\usepackage[normalem]{ulem}
\usepackage{units}
\usepackage{verbatim}

\usepackage[super]{nth}
\usepackage{lscape}
\usepackage{multirow}
\usepackage{hyperref}

\usepackage{graphicx}
\graphicspath{ {.} }

\usepackage{listings}
\usepackage{amssymb}

\usepackage{csquotes}

\usepackage{pdfpages}

\usepackage{url}
\makeatletter
\g@addto@macro{\UrlBreaks}{\UrlOrds}
\makeatother

\newcommand{\remove}[1]{}

\newcommand{\TOTAL}{\mathit{Total}}
\newcommand{\BRUTE}{\mathit{Brute}}
\newcommand{\ID}{\mathit{ID}}

\begin{document}

\title{Cross Layer Attacks and How to Use Them (for DNS Cache Poisoning, Device Tracking and More)}

\author{\IEEEauthorblockN{Amit Klein}
\IEEEauthorblockA{Bar-Ilan University, Israel\\
aksecurity@gmail.com}
}

\maketitle

\begin{abstract}
We analyze the prandom pseudo random number generator (PRNG) in use in the Linux kernel (which is the kernel of the Linux operating system, as well as of Android) and demonstrate that this PRNG is weak. The prandom PRNG is in use by many ``consumers'' in the Linux kernel. We focused on three consumers at the network level -- the UDP source port generation algorithm, the IPv6 flow label generation algorithm and the IPv4 ID generation algorithm. The flawed prandom PRNG is shared by all these consumers, which enables us to mount ``cross layer attacks'' against the Linux kernel. In these attacks, we infer the internal state of the prandom PRNG from one OSI layer, and use it to either predict the values of the PRNG employed by the other OSI layer, or to correlate it to an internal state of the PRNG inferred from the other protocol. 

Using this approach we can mount a very efficient DNS cache poisoning attack against Linux. We collect TCP/IPv6 flow label values, or UDP source ports, or TCP/IPv4 IP ID values, reconstruct the internal PRNG state, then predict an outbound DNS query UDP source port, which speeds up the attack by a factor of x3000 to x6000. This attack works remotely, but can also be mounted locally, across Linux users and across containers, and (depending on the stub resolver) can poison the cache with an arbitrary DNS record. Additionally, we can identify and track Linux and Android devices -- we collect TCP/IPv6 flow label values and/or UDP source port values and/or TCP/IPv4 ID fields, reconstruct the PRNG internal state and correlate this new state to previously extracted PRNG states to identify the same device.
\end{abstract}


\section{Introduction}
\subsection{DNS and DNS Cache Poisoning}
DNS (Domain Name System) \cite{rfc1035} is a set of protocols and conventions describing how an application can translate a host name into an IPv4/IPv6 network address. This process is called DNS {\em resolution}. In order to resolve a name into an address, the application uses a standard operating system API e.g. {\tt getaddrinfo()}, which delegates the query to a system-wide service called {\em stub resolver}. This local (on-machine) service in turn delegates the query to one of the name servers in the operating system's network configuration, e.g. an ISP/campus/enterprise name server, or a public name server such as Google's 8.8.8.8. This {\em recursive resolver} does the actual DNS resolution against the {\em authoritative} DNS servers that are responsible for sub-trees of the hierarchical DNS global database. Both the stub resolver and the recursive resolver may cache the DNS answer for better performance in subsequent resolution requests for the same host name.

DNS is fundamental to the operation of the Internet/web. For example, every non-numeric URL requires the browser to resolve the host name before a TCP/IP connection to the destination host can be initiated. Likewise, SMTP relies on DNS to find the network address of mail servers to which emails should be sent. Therefore, attacks that modify the resolution process, and specifically attacks that change existing DNS records in the cache of a stub/recursive resolver or introduce fake DNS records to the cache, can result in a severe compromise of the user's integrity and privacy. Our focus is on poisoning the cache of the Linux stub resolver.

The DNS protocol is implemented on top of UDP, which is a stateless protocol. In order to spoof a DNS answer, the attacker needs to know/guess all the UDP parameters in the UDP header of the genuine DNS answer, namely the source and destination network addresses, and the source and destination ports. We assume the attacker knows the destination network address, which is the address of the stub resolver, and the source network address, which is the address of the recursive name server used by the stub resolver. The attacker also knows the UDP source port for the DNS answer, which is 53 (the standard DNS port), and thus the only unknown is the destination port (nominally 16 bits, practically about 15 bits of entropy), which is randomly generated by the stub resolver's system.
At the DNS level, the attacker needs to know/guess the transaction ID DNS header field  (16 bits, abbreviated ``TXID''), which is randomly generated by the DNS stub resolver, and the DNS query itself, which the attacker can infer or influence. 
Thus, the attacker needs to predict/guess 31 bits (the UDP destination port, and the DNS TXID) in order to poison the cache of the stub resolver. Spoofing and sending $2^{31}$ DNS answers is almost impractical to carry out over today's Internet within a reasonable time frame, and therefore improvements to DNS cache poisoning techniques that can make them more practical are a topic of ongoing research.

\subsection{Device Tracking}
Browser-based tracking is a common way in which advertisers and surveillance agents identify users and track them across multiple browsing sessions and websites. As such, it is widespread in today's Internet/web. Web-based tracking can be done directly by websites, or by advertisements placed in websites. 

\subsection{Our Approach}
We analyze the {\tt prandom} PRNG, which is essentially a combination of 4 linear feedback shift registers, and show how to extract its internal state given a few PRNG readouts. 
For DNS cache poisoning, we obtain partial PRNG readouts by establishing multiple TCP/IPv6 connections to the target device, and observing the flow labels on the TCP packets sent by the device (on recent kernels, we can alternatively establish TCP/IPv4 connections and observe the IP ID values). This enables us to extract the internal PRNG state. Then we remotely force the device to emit a DNS query over UDP for a host name we control. Finally, we send a burst of spoofed DNS answers (going over all possible TXID values, with the UDP port predicted from the extracted PRNG state) containing an {\em arbitrary} DNS record. This record is cached by the stub resolver, which results in DNS cache poisoning. There is a PRNG instance per CPU logical core, hence it is possible that the UDP source port was generated by a PRNG different than the one generated the flow labels. However, by repeating this procedure a few times, we can gain a statistical certainty that the target's DNS cache would be poisoned with the arbitrary DNS record.

For device tracking, we use an HTML snippet which forces the browser running on the device to send TCP/IPv6, UDP, or TCP/IPv4 traffic to our website over multiple TCP/UDP connections. This allows us to collect sufficient amount of TCP/IPv6 flow labels, and/or UDP source ports and/or TCP/IPv4 IDs, to reconstruct the PRNG internal state of a single core. Since there is a PRNG instance per logical CPU core, we repeat this several times to extract the internal states of as many cores as possible. We then compare the states we extracted with the states of devices we already recorded, by virtually testing $N$ increments (where $N$ is proportional to the time duration in which devices are to be tracked, see Section~\ref{tracking}) of all recorded states and looking for a match with one of the presently extracted states. If a match is found, we can assume that it is the same device as the one recorded earlier. If no match is found, we record the extracted states as a new device.

\subsection{Advantages of Our Techniques}
\label{sec:advantages}
Our DNS cache poisoning technique is approximately 3000-6000 times faster than a brute force attack, making it very practical in the present day Internet connection bandwidth. It works completely remotely, i.e. it does not need any malicious software installed on the target machine. Unlike \cite{DBLP:journals/corr/abs-1205-4011,DV++,first-try-dns-cache-poisoning}, our technique does not rely on IPv4 fragmentation, which can be easily turned off. 

Our device tracking technique works across browsers, across the browsers' privacy mode, and across different networks, including IPv4-IPv6 and VPNs. It works even in the ``golden image'' scenario (a master disk image cloned to all desktops/laptops in an organization) and cannot be easily countered or turned off. It provides $\approx$57.7 bits of entropy, and is thus scalable. Finally, the tracking scope is from system startup to system shutdown/restart, hence providing a long-lasting identifier. We are not familiar with any other (un-fixed) device tracking technique that can fulfill all these requirements at the time of writing. However we note that the need for relatively long ``dwell time'' (time in which the browser needs to remain at the site) can hinder the applicability of this tracking technique in some use cases. 

\subsection{Our Contributions}
\label{sec:our-contributions}
Our contributions are threefold:
\begin{itemize}
    \item Analysis of the Linux {\tt prandom} PRNG and pinpointing its weaknesses. 
    \item Exploitation of the Linux {\tt prandom} PRNG weakness to mount a DNS cache poisoning attack against Linux hosts. Depending on the stub resolver software, this attack can introduce an {\em arbitrary} DNS record into the stub resolver cache. We ran DNS cache poisoning attacks over the Internet in various real-life scenarios, demonstrating successful attacks against multi-core hosts using two recursive resolvers taking 28.58s on average (transcontinental, effective bandwidth 30Mb/s) and 54.42s (transatlantic, 25Mb/s). A brute force attack under the same conditions takes roughly 2 days on average. Our attack is also almost 4.5 orders of magnitude more efficient in the number of spoofed packets sent.
    \item Exploitation of the Linux {\tt prandom} PRNG weakness to track Linux and Android devices. We demonstrated device tracking over the Internet with multiple diverse devices. We can use either UDP/IPv4, UDP/IPv6, TCP/IPv6, and -- in recent Linux kernels -- TCP/IPv4 traffic, thus crossing the IPv4/IPv6 gap. This demonstrates a {\em cross (OSI) layer attack} for Linux/Android. Table~\ref{tab:protocol-fields} summarizes the protocols and fields we can exploit. Specifically, for recent kernel clients, we can tie every combination of layer 3 and 4 protocols to all other combinations.
\end{itemize}

\begin{table}[]
\caption{Vulnerable Protocol Fields}
\label{tab:protocol-fields}
\begin{center}
\begin{tabular}{|l|l|l|}
\hline
 & TCP & UDP \\ \hline
IPv4 & \begin{tabular}[c]{@{}l@{}}IPv4 ID\\ (recent kernels)\end{tabular} & \begin{tabular}[c]{@{}l@{}}UDP Source Port\\ (UDP client)\end{tabular} \\ \hline
IPv6 & IPv6 Flow Label & \begin{tabular}[c]{@{}l@{}}UDP Source Port\\ (UDP client)\end{tabular} \\ \hline
\end{tabular}
\end{center}
\end{table}

\section{Related Work}
\subsection{Cross Layer Attacks}
\label{sec:related-cross-layer}
Cross layer attacks combine vulnerabilities across multiple network protocol layers to attack the target system, and/or exploit vulnerabilities in some network protocol layers to attack protocols in other layers. To date, only a handful of cross layer attacks are known \cite{cross-layer-MANET}, and all of them, to the best of our knowledge, involve the lower layers of the OSI model which are media specific, i.e. layers 1-2, below the IP stack. Some notable examples are \cite{lion-attack} which uses layer 1/2 to to degrade TCP bandwidth, and \cite{cross-layer-wireless} which inflicts a DoS condition on a wireless network using a cross layer attack on the low layers and the local routing layers. All these cross layer attacks require proximity, since they target the low network layers (which operate at the local network level). None of these attacks targets a more generic OSI layer such as layer 3 (IPv4/IPv6) and up, and none of them inflicts damage other than network performance degradation and local routing modifications. Our cross layer attacks work at the IP layer and up, and therefore do not require proximity, and their security impact is different and more severe: DNS cache poisoning, device tracking across IPv4/IPv6 networks, and more. 

\subsection{DNS cache poisoning}
In general, DNS cache poisoning attacks either try to guess/predict the 16 bits DNS transaction ID (TXID) header field and the UDP source port of the query (16 bits field in the UDP header, typically 14-16 bits of entropy), or, using fragmentation, to guess/predict the 16 bits IPv4 ID header field of the UDP datagram.
A brute force attack, enumerating over both TXID and UDP source ports, is documented in \cite{Gbyte-attack}. Brute force attacks are much inferior to our attacks, as can be seen in Table~\ref{tab:poisoning-results}. 
An attack against Android's randomness of TXID and UDP source port was published in 2012 \cite{weak-randomness-android}, but these issues were since addressed.
A port exhaustion attack against Linux and Windows, leading to DNS cache poisoning was described in 2011 \cite{port-exhaustion-hay}, however this attack requires running Java in the browser, which was since disabled in all major Linux browsers. Another port exhaustion attack against various operating systems, including Linux, is described in \cite{port-exhaustion}, but this attack requires a local malware running on the target machine, which is a very restrictive requirement. A cache poisoning attack against local DNS forwarders, using a cooperative agent on the same LAN is described in \cite{255314}. Our DNS cache poisoning attack is completely remote, and does not rely on a cooperative agent on the target machine or on its LAN. 
Alternatively, DNS cache poisoning attacks can exploit UDP fragmentation by forcing fragmentation of the DNS response into 2+ UDP fragments, and spoofing the \nth{2} fragment by keeping the same UDP checksum as the original fragment, and enumerating over the IP ID. Such attacks are described in \cite{DBLP:journals/corr/abs-1205-4011,DV++}, with an improvement (predicting the IP ID value in the DNS answer UDP datagram) in \cite{first-try-dns-cache-poisoning}. However, IP fragmentation in DNS answers can be systematically and easily eliminated in today's Internet, as proposed in IETF draft standards \cite{I-D.fujiwara-dnsop-fragment-attack,I-D.fujiwara-dnsop-avoid-fragmentation}, e.g. disabling fragmentation of DNS answers in authoritative name servers and dropping fragmented DNS answers in DNS resolvers -- the same can be applied to DNS resolvers and stub resolvers, respectively. 

\subsection{Device Tracking}
\label{sec:related-work-tracking}
A fairly recent review (early 2019) of web-based tracking methods can be found in \cite{KP18}. It evaluates the known tracking techniques against the challenges of ``golden image'' and the browser privacy mode, as well as estimates the coverage of the techniques in question. They find no tracking technique with adequate performance in all three categories. That paper also describes a DNS cache-based technique. However, this technique does not survive network switching (moving the device from one network to another). 

Additional techniques developed since are \cite{KP19-usenix} which describes an IPv4 ID based device tracking (for Linux/Android, this only applies to UDP/IPv4), and \cite{flowlabel} which describes an IPv6 flow label based device tracking (for Linux/Android, this only applies to UDP/IPv6). In both cases, while the approach is somewhat similar, there is a major difference, which is that in the above cases, a cryptographic {\em key} is extracted, whereas in our case, the situation is more challenging since there is no static key -- rather, we rely on our ability to extract and associate PRNG state values in different times. The attacked fields in both cases do not overlap with our attacked fields -- for example, \cite{flowlabel} exploits the IPv6 flow label field in {\em UDP} traffic, whereas we exploit the IPv6 flow label in {\em TCP} traffic (where the flow label is generated in a completely different manner). Also, the underlying flaws in both cases were addressed by the respective kernel makers in 2019, and are therefore no longer in effect. 
It should be noted that \cite{flowlabel} only provides 32 bits of entropy in its device ID, whereas our technique provides $\approx$57.7 bits. Therefore, for large scale tracking, our technique is superior. For instance, when tracking 100,000,000 devices, \cite{flowlabel} yields more than 1,000,000 expected device ID collisions (hence, about 2\% of the devices will have non-unique IDs), whereas our technique is very likely not to yield any collisions. Finally, both techniques do not cross the IPv4/IPv6 gap, i.e. they cannot track a device connected to an IPv4 network when it moves to an IPv6 network, and vice versa. For example, \cite{flowlabel} is limited to IPv6 devices and networks. In contrast, our technique survives network switching, including IPv4/IPv6 network switching.

Kohno et al. \cite{tcp-clockskew} demonstrate device identification based on the device clock skew, observed in the TCP timestamp field. However, the entropy obtained from this technique is too low (4.87-6.41 bits \cite[Table 2]{tcp-clockskew}) to serve as a standalone device ID. This technique was later improved \cite{hot-or-not} to vary the CPU temperature and thus modulate the CPU clock frequency, but this only obtains additional ``2–8 bits per hour'' \cite[Section 4]{hot-or-not} and is therefore impractical for real-world tracking.

A recent survey of sensor-based tracking device identification techniques is presented in \cite[Table 1]{sensors}. As can be seen, none of the techniques surveyed is both practical and performant. The new technique introduced in that paper also yields results \cite[Table 5]{sensors} which make it impractical when the device is not in a ``flat static case''. 

A tracking technique specific to the Apple Safari browser is described in \cite[Section~2.3]{janc2020information}. However, this technique only applies to Safari, whose market share is quite limited, especially w.r.t. Linux and Android devices.

\section{The Linux kernel {\tt prandom} Pseudo Random Number Generator}
\subsection{Implementation}
We examine the {\tt prandom} pseudo-random number generator (PRNG) implemented in the Linux kernel since version 3.13-rc1. This PRNG has per-core states. The PRNG state used when invoked is the one that belongs to the {\em logical} core on which the kernel thread presently executes. Each core-state consists of 4 linear feedback shift registers (LFSRs) denoted as $S_1, S_2, S_3, S_4$, whose bit lengths are $k_1=31, k_2=29, k_3=28, k_4=25$ respectively, and whose taps are fixed, implemented as 4 32-bit unsigned integers. Each $S_i$ is advanced a fixed number $s_i$ of steps in every PRNG invocation. Finally, the 32 bit states are XOR-ed together to form the 32-bit PRNG output {\tt prandom\_u32()}.
A helper function {\tt prandom\_u32\_max($m$)} provides a random integer $r$, $0 \leq r < m$ by calculating $r=\lfloor\frac{m \cdot \tt prandom\_u32()}{2^{32}}\rfloor$.
We see therefore that each LFSR in the PRNG core-state is advanced via a known linear transformation conforming to $s_i$ LFSR steps, and the PRNG output is a linear combination of the LFSRs: $S_1 \oplus S_2 \oplus S_3 \oplus S_4$. For simplicity, we consider those $s_i$ basic steps as a single LFSR (or $S_i$) step. The combined linear transformation matrix rank over the vector space $GF(2)^{128}$ is $k_1+k_2+k_3+k_4=113$.
While a PRNG instance is completely linear, in practice the analysis of the {\em system} is complicated by the existence of multiple PRNG instances, multi-threading, frequent re-seeding, and interrupts, as described below.

\subsection{Seeding and Re-Seeding}
\label{sec:re-seeding}
All core states are fully seeded at system startup, and thereafter they are re-seeded simultaneously every 40-80 seconds (the exact time interval is randomly chosen right after re-seeding). Re-seeding involves obtaining a cryptographically strong random 32-bit quantity from another kernel PRNG, and XOR-ing this quantity into $S_1$ of each core state.

\subsection{Interrupt-Safety}
\label{sec:multithreading}
The PRNG code is thread-safe (i.e. no two threads running simultaneously can access the same core-state), but it is {\em not} interrupt-safe. A thread can start executing the PRNG algorithm, then an interrupt can be triggered, which causes an interrupt handler to start running, and the latter may run the PRNG logic for the same PRNG core-state. Each LFSR is updated atomically because it is word-aligned, but the complete state of the PRNG is {\em not} updated atomically, and therefore having a thread and an interrupt handler updating the same core-state can affect the functionality of the PRNG.

\subsection{Consumers}
{\tt prandom\_u32()} is only exposed to the kernel code, and cannot be directly invoked from userspace. There are 108 Linux kernel v5.5.6 files in which {\tt prandom\_u32()} is invoked,\footnote{\url{https://elixir.bootlin.com/linux/v5.5.6/ident/prandom_u32}} and 26 Linux kernel files in which {\tt prandom\_u32\_max()} is invoked.\footnote{\url{https://elixir.bootlin.com/linux/v5.5.6/ident/prandom_u32_max}} These consumers are diverse, and include file-systems, drivers and network protocols.
Since we are interested in remote exploitation attacks, the network consumers are the most relevant. Particularly, we focus on the invocations that are in the TCP and UDP {\tt connect()} IPv6 flow label generation, the UDP source port generation and (in recent Linux kernels) the IPv4 ID generation for TCP.
\subsubsection{TCP and UDP {\tt connect()} (over IPv6) flow label generation}
Starting with Linux kernel version 4.3-rc1, the default Linux flow label generation algorithm for TCP, and for UDP {\tt connect()} (over IPv6), invokes {\tt prandom\_u32()} once, and generates the flow label by swapping the two 16-bit halves of the result\footnote{swapping was introduced in kernel version 4.18-rc6, and back-ported to 4.17.10, 4.14.58, 4.9.115 and 4.4.144.} and then extracting the 20 least significant bits of the result to form the 20 bits IPv6 flow label. 
\subsubsection{UDP Source Port Generation}
The Linux UDP source port generation algorithm invokes the PRNG once, and generates the UDP source port $P$ ($L \leq P \leq H$, where $L$ is the lowest port number allowed, and $H$ -- the highest) as $P={\tt prandom\_u32\_max}(H-L+1)+L$. The values of $(L,H)$ are hard-wired and depend on the kernel version: for kernel version 4.2-rc1 and above, they are $(32768,60999)$, for earlier kernels they are $(32768,61000)$, and for some Android kernels they are $(37000,50000)$.
\subsubsection{TCP/IPv4 ID generation}
Starting with version 5.4-rc6, the Linux kernel generates the initial IPv4 ID for a TCP connection by invoking {\tt prandom\_u32()} and extracting the least significant 16 bits. This algorithm was back-ported to kernel versions 5.3.10, 4.19.83, 4.14.153, 4.9.200, 4.4.200 and 3.16.80. We define {\em recent Linux kernels} as any Linux kernel in those branches at least as recent as the listed versions, and of course version 5.4 and later. The initial IPv4 ID for TCP is generated with the SYN packet for clients, and with the reception of the final handshake ACK packet for servers which will use it for the first packet sent after the handshake.
\subsubsection{Additional Noteworthy Consumers}
\begin{itemize}
    \item UDP {\tt connect()} consumes one PRNG value even in IPv4 (two values for recent kernels).
    \item Any UDP packet over IPv4 may consume one PRNG value, during the IP ID generation. The IP ID generation algorithm invokes the PRNG once, if the time since the last use of the present IP ID bucket is at least one ``jiffie'' tick (in Linux - 4ms) older than the present time.
\end{itemize}

\section{Analysis of the {\tt prandom} PRNG}
\subsection{Obtaining the State of the PRNG on a Single-Core Device}
\label{sec:cryptanalysis}
Clearly, if we know that a sequence of PRNG outputs originates from a single core, and we know the usage pattern, namely how many (if at all) ``hidden'' PRNG consumption instances occurred between each observed PRNG output value, then obtaining the state of the PRNG e.g. at the end of the sequence, amounts to solving linear equations over 113 unknowns in $GF(2)$. This can be done easily, if we define, for $S_i$ ($1 \leq i \leq 4$) $k_i$ unknowns in $GF(2)$ and we keep track of the LFSR state as a function of these unknowns, given that $S_i$ is advanced by applying $s_i$ times the basic $i$-th LFSR step. Therefore, each bit of the {\tt prandom} PRNG output yields a single linear equation over $GF(2)$ on the set of $k_1+k_2+k_3+k_4=113$ unknowns. Thus, 113 bits of {\tt prandom}, all coming from the same core, suffice to reconstruct its internal core state, and generate future values, using standard linear equation solving. This involves $113^3$ bit operations, which, on a modern machine, run in less than 1ms.

In the case of IPv6 flow label, we observe the least significant 20 bits of each PRNG output (after swapping). Since these bits are directly extracted from the PRNG output, we can use them as-is, thus we need at least 6 flow label values (120 bits) in order to solve the linear equations. Note that \cite{flowlabel} reports no evidence of flow label manipulation en-route. 

The case of TCP IPv4 ID (for recent Linux kernels) is similar to the above, with the 16 least significant bits of each PRNG output forming the observed ID field. Thus, 8 values of TCP/IPv4 IDs suffice to solve the linear equations. Recent research \cite{KP19-usenix} showed that 92\% of the networks observed did not modify TCP/IPv4 ID values.

The case of UDP source ports is only relevant for device tracking, and is more complicated. Given a source port value $P$ which we observe, we want to formulate $k$ (the choice of $k$ is described below) linear equations on the state of the PRNG that generated $P$.

Assuming we know/guess $(L,H)$, we have $P-L=\lfloor\frac{(H-L+1) \cdot x}{2^{32}}\rfloor$ where $x$ is the original 32-bit PRNG output. It follows that $\lceil \frac{2^{32}(P-L)}{H-L+1} \rceil \leq x \leq \lfloor \frac{2^{32}(P-L+1)}{H-L+1} \rfloor$. This defines, for a given $P$, an interval $X$ of the possible $x$ values, whose size is approximately $\frac{2^{32}}{H-L+1}$ (more accurately, it is either $\lfloor \frac{2^{32}}{H-L+1} \rfloor$ or $\lceil \frac{2^{32}}{H-L+1} \rceil$). We are interested in the $k$ ($0<k < \log_2(H-L+1)$) most significant bits of the values in this interval. Let us define $2^k$ intervals $I_i=[i \cdot 2^{32-k},(i+1)\cdot 2^{32-k})$, where each $I_i$ conforms to the $k$ most significant bits being $i$. By the definition of $k$, $2^{32-k} > \frac{2^{32}}{H-L+1}$, hence $2^{32-k} \geq \lceil \frac{2^{32}}{H-L+1} \rceil$ i.e. $|I_i| \geq |X|$. Thus, for $i^*=\lfloor \frac{\lceil \frac{2^{32}(P-L)}{H-L+1} \rceil}{2^{32-k}} \rfloor$, $X \subseteq I_{i^*} \cup I_{i^*+1}$.

We now calculate the probability $p_k$ of $X \subseteq I_{i^*}$. A complete inclusion happens if the starting point of $X$ is within $2^{32-k}-\frac{2^{32}}{H-L+1}$ of the starting point of $I_{i^*}$. Therefore, the probability of complete inclusion is $p_k=\frac{2^{32-k}-\frac{2^{32}}{H-L+1}}{2^{32-k}}=1-\frac{2^k}{H-L+1}$. 

Hence, we can extract a single value $i^*$ for the most significant $k$ bits of $x={\tt prandom\_u32()}$ from $P$, with probability $p_k$. In case $X \nsubseteq I_{i^*}$, we know that $X \subseteq I_{i^*} \cup I_{i^*+1}$. In such case, we can take one of the following approaches:
\begin{itemize}
    \item Discard the sample.
    \item ``Guess'' the most likely value for the $k$ most significant bits.
    \item Enumerate over the two possible $k$ most significant bits in the interval: $i^*$ and $(i^*+1)$.
\end{itemize}
In the ``guess'' case, we choose the $k$ most significant bits that cover most of the $X$ interval, i.e. $i^*$ if $|I_{i^*} \cap X| \geq |I_{i^*+1} \cap X|$, $i^*+1$ otherwise, hence we stand a chance $\ge 0.5$ that the guess is correct. Therefore, the overall probability $p^G_k$ for extracting a correct $k$ most significant bits using the ``guess'' approach satisfies $p^G_k \ge p_k+0.5(1-p_k)=\frac{1+p_k}{2}=1-\frac{2^{k-1}}{H-L+1}$. This is the approach we took in our experiments.

Since each sample $P$ yields $k$ most significant bits of the PRNG output, i.e. $k$ linear equations, we need at least $\lceil \frac{113}{k} \rceil$ samples to find the PRNG core-state. 

The probability of extracting all the necessary linear equations correctly, i.e. the probability of $\lceil \frac{113}{k} \rceil$ samples to have a correct $k$ significant bits value in all the $x$ values is bounded from below by ${(p^G_k)}^{\lceil \frac{113}{k} \rceil}$. When $\frac{2^{k-1}}{H-L+1} \ll 1$, this is approximately $e^{-\frac{2^{k-1}}{H-L+1}\lceil \frac{113}{k} \rceil} $. For example, with $k=6$ and $L=37000, H=50000$ (worst case scenario in terms of $L,H$) we get a lower bound of 0.954. 

Lower $k$ values increase the probability of the above technique to successfully extract a correct set of equations, but at the same time decrease the probability of the complete set to represent samples in predictable intervals from a single core, since lower $k$ means longer sequences needed. In our experiments, we tried multiple values of $k$ for the same sequence of samples ($8 \leq k \leq 13$, which we found experimentally to be optimal), thereby increasing the overall probability to succeed.

Note: the above discussion assumes that the attacker can observe the UDP source port used by the target device. This is not always the case for UDP/IPv4 source ports, as there may be network devices (NATs, firewalls, gateways) that modify the UDP source port. This is less likely for IPv6 networks as NAT is usually not needed there. In Section~\ref{sec:tracking-experiment} we report that 59\% of the IPv4 networks we surveyed preserve UDP source ports. Since our calculation above does not use the least significant bits of the port numbers, we can use these bits to verify the correctness of the core-state extraction. A match of all $\lceil \frac{113}{k} \rceil$ values of the actual least significant $\log_2(H-L+1)-k$ bits of the $P$ values with the predicted ones from the PRNG core-state extraction statistically verifies that the port numbers are generated by the Linux kernel, and are not modified en-route. Failure to match the least significant bits may indicate a UDP source port modification en-route.

\subsection{Multi-Core Devices}
The above analysis assumes that the PRNG outputs originate from a single core. When the device has multiple cores, this condition does not automatically hold. However, we noticed that when the PRNG is invoked in a very short time interval, there is a good chance that the sequence will in fact originate from the same core. Therefore, it is crucial that our interaction with the target device enables us to extract PRNG values generated in a rapid succession (burst). 

Another complication in multi-core devices is that knowing a PRNG state for one core does not guarantee the ability to predict the PRNG value even after few hundred milliseconds, due to the fact that such PRNG invocation may originate from a different core.

\subsection{Observation Linking and Overcoming Re-Seeding}
\label{tracking}
In this sub-section, we describe how we can link two observed PRNG states of the same device, even across re-seeding events. This analysis is needed for device tracking. A reader interested only in DNS cache poisoning can skip this sub-section.

As noted in Section~\ref{sec:re-seeding}, re-seeding the PRNG core-state only modifies $S_1$. Therefore, if we know the states $S_2, S_3, S_4$ of a core-state at time $t_1$, and then at a later time $t_2$ we get another readout of $S_2, S_3, S_4$ from the same core-state, it is guaranteed that all  $S_2, S_3, S_4$ states advanced the same number of steps from $t=t_1$ to get to their values at $t=t_2$. This lends itself to a linking (of observations of the same PRNG in different times) technique as follows: assume we are dealing only with single core devices, and assume we need to support up to $R$ devices, and link an observation to the correct device when the $S_i$ LFSRs advance up to $N$ steps (there is a strong correlation between the number of PRNG steps and the time elapsed, hence this assumption translates to the time duration we want to support for observation linking). We also assume $R \cdot N \ll 2^{k_2+k_3+k_4}=2^{82}$. 

For an examined device, we extract $S_2, S_3, S_4$ as described above. We apply the baby-step giant-step technique \cite{1971-shanks} to find a previous observation of the device, if such exists. This is elaborated in Appendix~\ref{app:baby-step-giant-step}, and involves a table $T$ that maintains $\sqrt{N}$ ``future'' states (in intervals of $\sqrt{N}$) of previously observed devices. Appendix~\ref{app:baby-step-giant-step} also explains how to extend the technique to support multi-core devices, and analyzes the device ID entropy.

We can safely assume, based on measurements from several devices reported in Appendix~\ref{app:consumption-rate}, that in a single day, an active Android device consumes less than 100,000 PRNG outputs. Therefore, if we want to support observation linking across up to 100 days, we can set $N=10^7$. If we want to support up to a million devices, then $R=10^6$. Thus we have $R \cdot N=10^{13}=2^{43.185} \ll 2^{82}$, as required. From Appendix~\ref{app:baby-step-giant-step}, we have lookup/insert time $\approx 3000$ operations, the table ($T$) size is $\approx 3 \times 10^9$ entries, and the device ID entropy for this choice of $N$ is 57.747 bits.

A consequence of the non-interrupt-safe implementation of the PRNG is that on some devices, we may encounter relative ``drift'' in the LFSR states. This is treated in Appendix~\ref{app:drift}.

\section{DNS Cache Poisoning (Linux only)}
\label{cache-poisoning-general}
Our DNS cache poisoning attacks target Linux hosts, and since for Linux, only versions $\ge4.4$ are supported, we can assume $L=32768, H=60999$. Caching DNS stub resolvers are either {\em record-based} or {\em query-based} (see \cite{KP18}). A record-based caching stub resolver caches the individual records extracted from DNS answers in a ``flat'' database, so all such records are accessible for a later cache lookup, regardless of the query. In contrast, a query-based caching stub resolver caches the entire answer for the query atomically, so it is only consulted in a cache lookup for the exact same query.

Our DNS cache poisoning attack works best against record-based caching DNS stub resolvers such as Ubuntu's default stub resolver  -- Systemd-Resolved. Therefore we focus on record-based caching DNS stub resolvers, and specifically on Systemd-Resolved. We stress that the attack can be mounted against query-based DNS stub resolvers (e.g. glibc NSCD daemon -- Appendix~\ref{app:attacking-query-based}, and glibc in-process stub resolver -- Appendix~\ref{app:attacking-cacheless}), albeit with reduced efficiently and coverage. However, the attack is not applicable to stub resolvers which do not defer the UDP source port selection to the operating system, e.g. Android's stub reslover -- bionic libc.

It should be noted that the impact of DNS cache poisoning extends beyond the service/application abused to achieve it (SMTP MTA in our example). Once the attack succeeds, {\em any} service/application on the target machine that resolves the poisoned DNS name will get the poisoned record from the cache, as long as it is not stale and not evicted. See a list of such examples in Appendix~\ref{app:dns-cache-poisoning-impact}.

\begin{figure}[h]
\centering
\caption{DNS Cache Poisoning Attack Entities}
\includegraphics[trim=7cm 1cm 8.7cm 1cm, clip=true, totalheight=0.32\textheight]{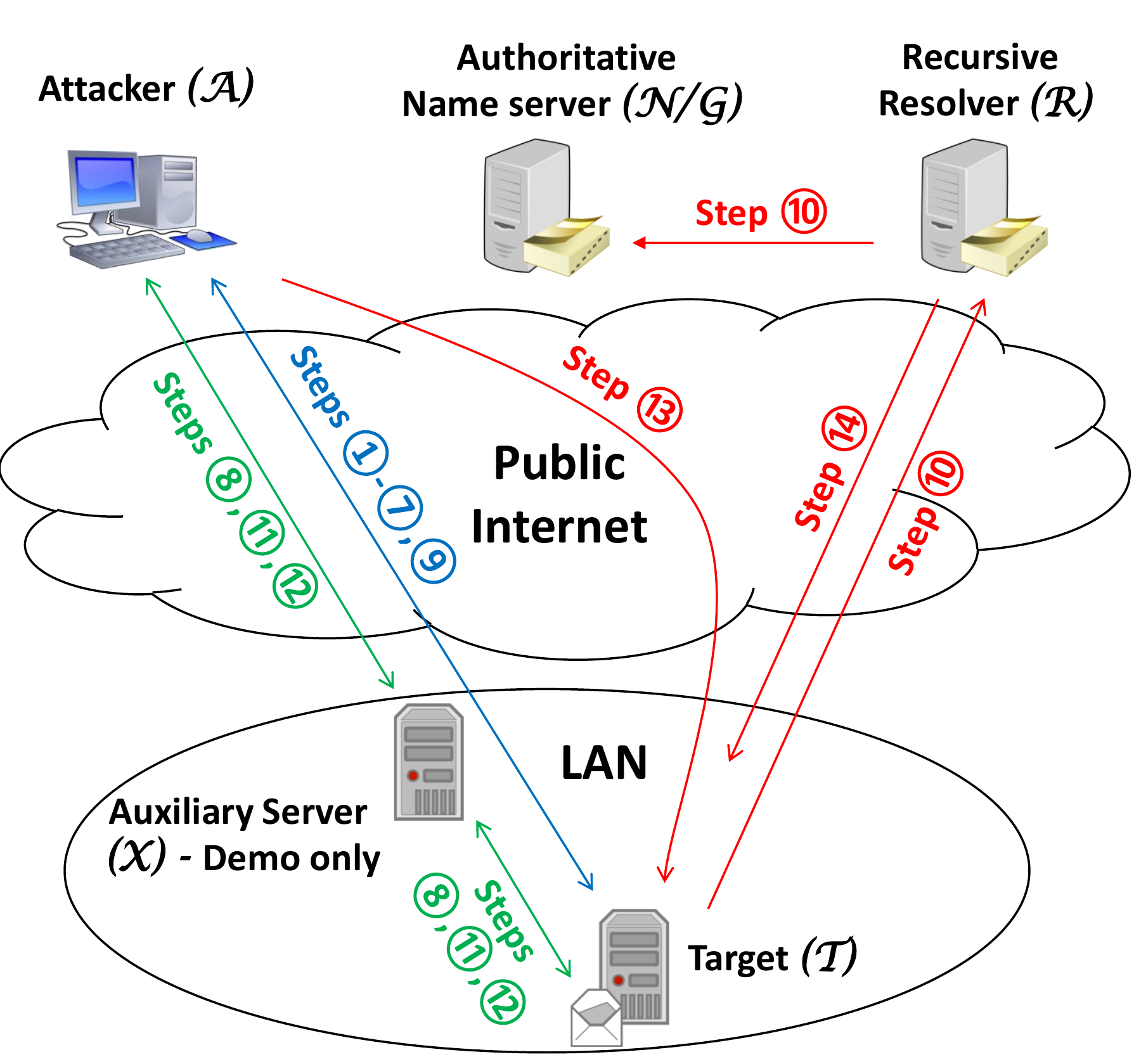}
\label{fig:topo}
\end{figure}

We denote by the calligraphic letters $\mathcal{A}, \mathcal{T}, \mathcal{R}, \mathcal{N}, \mathcal{G}$ and $\mathcal{X}$ the attacker machine, the target machine, the target's recursive resolver, the attacker's authoritative name server (when the recursive resolver employs record-based caching), a genuine authoritative name server (when the recursive resolver employs query-based caching) and an auxiliary machine in the target's LAN (for demonstration purpose only; when IPv6 encapsulation is needed), respectively. Fig.~\ref{fig:topo} is a schematic view of the entities involved. 

We denote round trip time (RTT) between two such entities using the double arrow symbol, e.g. $t_{\mathcal{A}\rightleftarrows\mathcal{T}}$ is the RTT between the attacker and the target machine.
We denote the recursive DNS resolver timeout as $t_O$, and we denote the number of configured IP addresses for recursive resolvers in the stub resolver of $\mathcal{T}$ as $|\mathcal{R}|$.

\subsection{Attacking a Record-Based Caching DNS Stub Resolver}
\label{sec:dns-attack-basic}
The DNS cache poisoning attack against a Linux target running a record-based caching DNS stub resolver consists of two steps. 

In Phase (a), the attacker learns the core-state of the {\tt prandom} PRNG, for one of the target's CPU cores. The core-state can be extracted by collecting a burst of UDP packets (with different UDP source ports) generated by the target system, or a burst of TCP packets over IPv6 (in different TCP circuits, hence different flow labels) generated by the target system, or (in recent kernels) a burst of TCP/IPv4 packets (in different TCP circuits). How an attacker obtains such packets is scenario dependent. Our example (and the system which we demonstrated the attack on) is a simple and realistic scenario in which the target is an SMTP server, serving incoming requests from the Internet over IPv6. In such case, the attacker can send multiple TCP SYN packets, and receive SYN+ACK responses that contain IPv6 TCP flow label values generated in a short period of time.

As described in Section~\ref{sec:cryptanalysis}, given a series of UDP source port numbers, or TCP/IPv6 flow label values, or TCP/IPv4 IPID values, it is possible to extract the core-state of the core that generated them, assuming they were all generated by the same core, which is very likely for a short burst of packets.

In Phase (b), the attacker predicts the UDP source port that will be used with the next DNS query, assuming it will be generated by the same core as in Phase (a). The predicted UDP source port of the DNS query is not necessarily generated from the PRNG core-state obtained at the end of Phase (a), as additional hidden PRNG consumption may occur prior to the forming of the DNS query. However, the attacker can predict, with sufficiently high probability, the offset in PRNG steps from the core-state obtained in Phase (a), of the PRNG state used to generate the UDP source port for the DNS query. This phase makes the attack so attractive, since predicting the UDP source port of the DNS query with high likelihood (e.g. probability of $\approx \frac{1}{11}$) yields roughly $\times 2500$ (over 3 orders of magnitude) gain over the brute force approach (which needs to go over all possible $H-L+1=28232$ values for UDP source port). We will describe an improvement below that can drive the gain up to almost $\times 6000$ in some cases.

\begin{table}[]
\caption{Resolver Software/Service Timeout}
\label{tab:timeout}
\begin{center}
\begin{tabular}{|l|l|r|}
\hline
\begin{tabular}[c]{@{}l@{}}Resolver\end{tabular} & Type & $t_O$ {[}s{]} \\ \hline
Google DNS (8.8.8.8) & Service & 1-2 \\ \hline
IBM Quad9 (9.9.9.9) & Service & 1.5 \\ \hline
Yandex.DNS (77.88.8.8)& Service & 2.5 \\ \hline
Cloudflare (1.1.1.1) & Service & 4 \\ \hline
OpenDNS (208.67.222.222) & Service & 6 \\ \hline
\begin{tabular}[c]{@{}l@{}}Microsoft DNS 6.2\\ (Windows Server 2012)\end{tabular} & Software & 7-8 \\ \hline
BIND 9.11.1 & Software & 10 \\ \hline
Unbound 1.6.0 & Software & 15 \\ \hline
\end{tabular}
\end{center}
\end{table}

The attacker forces the target machine to emit a DNS query to the attacker's domain. For example, an SMTP server which supports DKIM \cite{rfc6376}, when given a DKIM header, has to retrieve the DKIM public key for the sender domain via a DNS query; a similar approach using SPF is used in \cite{SPF-trigger}. The attacker then sends a burst of spoofed DNS answers for the predicted query -- enumerating over as many combinations of the possible 65536 DNS TXID values, and possible DNS recursive resolvers used by the target machine (typically 1-2 IP addresses) as the attack window allows. This window depends on the attacker's bandwidth and $t_O$. If indeed the attacker guessed the recursive server correctly, and if the core used for generating the DNS query source port was the same as the one that emitted the UDP source ports or TCP/IPv6 flow labels in Phase (a), and if the attacker's spoofed answer with the correct TXID value arrived to the target before the genuine answer from the recursive resolver, then the attacker's answer would be accepted by the stub resolver, and its content would be cached. 

{\em Poisoning the cache with arbitrary records:} in Systemd-Resolved, the answer records are cached regardless of the original query, overriding already-cached DNS records for the same names if they exist. Thus a successfully spoofed DNS answer can achieve DNS cache poisoning with {\em arbitrary} DNS records. However, caching arbitrary records is not a necessary condition -- it is possible to mount an attack against a stricter stub resolver, that accepts and caches only those records that are required to construct the answer for the query at hand, with a similar outcome.\footnote{This can be done using a CNAME chain: {\tt x.attacker.site CNAME www.google.com}, 
{\tt www.google.com CNAME y.attacker.site}, 
{\tt y.attacker.site TXT "foo"} -- this valid answer forces the stub resolver to cache an arbitrary CNAME record, which is practically as effective as caching an arbitrary record.} In both cases, the {\em payload} can contain multiple DNS records, up to the packet size limit. Henceforth, we use one record in order to minimize the packet size and thus the bandwidth we consume.

The attacker can significantly increase the window of opportunity for the spoofed answers to be accepted by the stub resolver, by simply having the authoritative domain server (under the attacker's control) refrain from answering queries from any recursive DNS resolver for the domain at hand. This can buy the attacker at least 1 second (the minimal timeout in Table~\ref{tab:timeout}) to send spoofed packets before the recursive server sends its own {\tt SERVFAIL} answer.

We now describe an improvement to the above scheme that reduces the attack time and the amount of spoofed packets needed. After the DNS query, we add a ``post-DNS'' probe (in the flow label case -- an additional TCP SYN packet). The DNS query thus becomes ``sandwitched'' between the pre-DNS probes and the post-DNS probe. Once the response for the post-DNS query arrives, the attacker can try to match its flow label to the PRNG states which follow the last known state from the pre-DNS probes.
Heuristically, a match for the post-DNS probe is highly correlated to a match in the UDP source port for the DNS query, since it means that the SMTP process is still running on the same core whose PRNG state was extracted. The time and packet savings come from having the attacker send the spoofed DNS answers only if a match for the post-DNS probe is found. When a match is not found, the attacker can start a new iteration immediately.

\begin{figure}[h]
\centering
\caption{DNS Cache Poisoning Attack Sequence}
\includegraphics[trim=6cm 0cm 5.6cm 0cm, clip=true, totalheight=0.32\textheight]{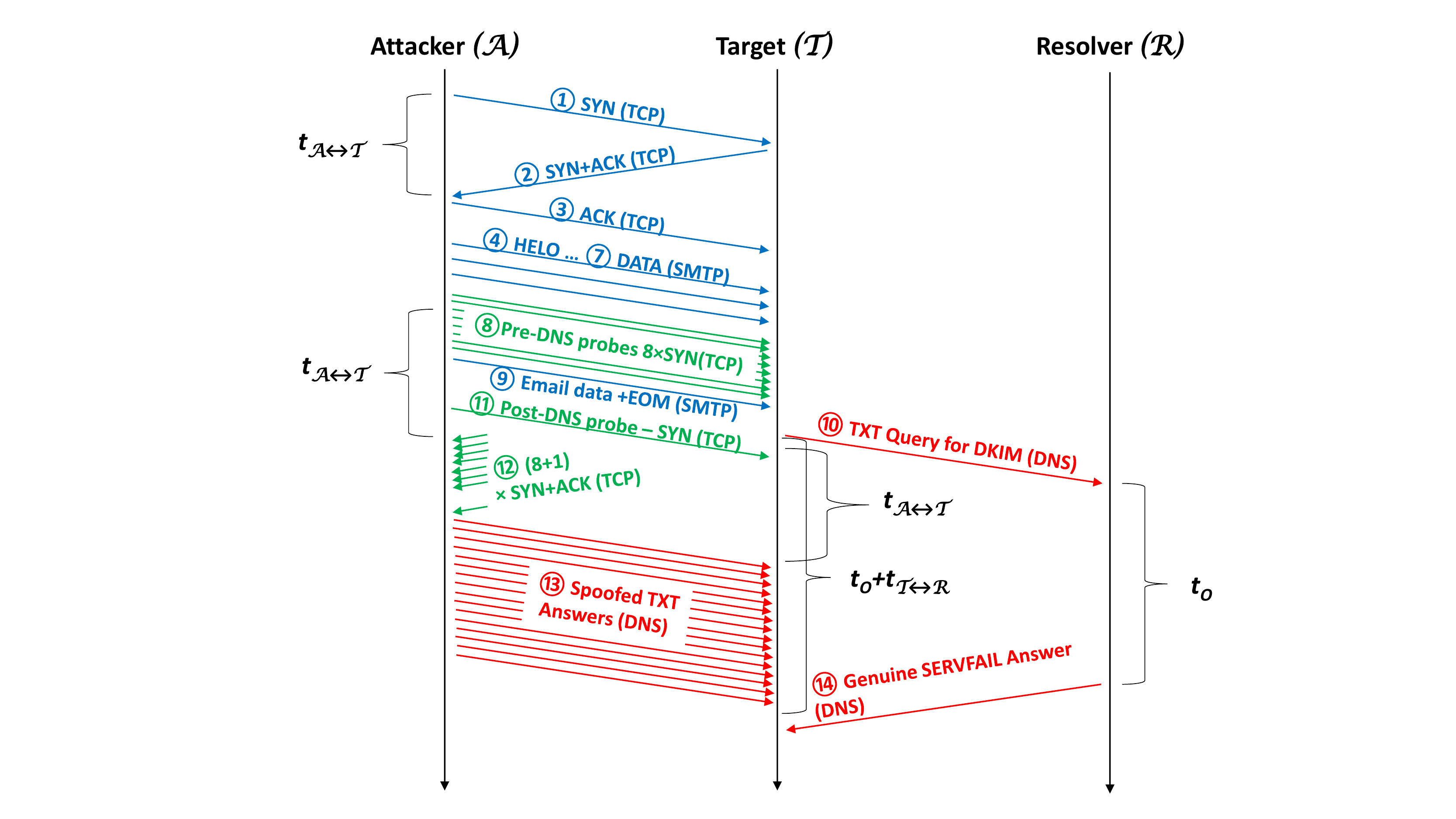}
\label{fig:msc}
\end{figure}

A message sequence chart describing the attack in more detail can be found in Fig.~\ref{fig:msc}. The blue messages mark the ``main'' SMTP session, which is used to force the target to emit a DNS query. The green messages are probing activities used to obtain the PRNG state, and the red messages mark the DNS query and the spoofed and genuine DNS responses.
The individual steps are detailed in Section~\ref{sec:payload} and Section~\ref{sec:poisoning}. As can be seen in Fig.~\ref{fig:msc}, the total time the attacker spends before sending the attack burst (step 13) is $2 \cdot t_{\mathcal{A}\rightleftarrows\mathcal{T}}+t_\Delta$, where $t_\Delta$ denotes the additional time consumed due to delays introduced by the attacker between steps. 
The attack step introduces additional $t_O+t_{\mathcal{T}\rightleftarrows\mathcal{R}}-t_{\mathcal{A}\rightleftarrows\mathcal{T}}$ time.

Of course, the attacker is free to reiterate the attack until DNS cache poisoning is achieved.

The above SMTP-based scenario is merely an example as well as the use case we explore in our experiments. Other attacks, e.g. against a forward/transparent HTTP proxy, are also possible.

\subsection{IP Source Address Spoofing}
\label{sec:source-address-spoofing}
The ability to send the target IP packets whose source address is spoofed (to have the IP address of the recursive resolver) is crucial for the DNS cache poisoning attack.

Some ISPs/networks employ filtering or NAT on their customer networks in such way that spoofed packets pertaining to originate outside the customer network are either dropped or modified to retain the genuine source IP address. According to \cite[Fig. 4]{Spoofer}, 11.9\%-30.5\% of the ASes do not employ such filtering, hence can be used by the attacker to send spoofed packets from. Naturally, the attacker is free to choose such a network, thus having enough non-filtering networks guarantees that the attacker can mount the attack.

As for the target's network -- some networks employ ingress filtering (also known as Source Address Validation -- SAV), which can identify a spoofed source address when the packet arrives form an interface which that does not serve its network. In its simplest and probably most popular form, an ISP may drop inbound packets from the Internet pertaining to originate from the ISP network. This practice neuters a DNS cache poisoning attack when the recursive resolver belongs the ISP or the ISP's customer. However, for public DNS services, this practice is less likely to interfere with the attack, since it requires an intermediate network in {\em every} route from the attacker to the target to carefully map and enforce the validity of the origin address in all inbound packets.

Recently, \cite{SAV} and \cite{SMap} measured inbound SAV, and discovered that it is {\em not} employed by 
48.93\%-87.4\% and 47.56\% of ASes, respectively. \cite{SAV} also reports that 63.98\%-76.28\% of the class C networks surveyed do not employ inbound SAV.

We sampled Internet networks/hosts and measured the portion that use a public DNS service for their recursive resolver. Based on our experiment described in Appendix~\ref{app:public-dns-use}, we estimate that 24\%-37\% of the hosts (depending on the organization size) use public DNS services. This makes the public DNS service use case alone a major exposure surface. As part of this experiment, we also describe how to identify a host that uses a public DNS service.

\subsection{UDP Source Port Preservation}
\label{sec:UDP-source-port-preservation}
For our UDP source port prediction technique to be effective, the UDP source port of the DNS query, which is generated by the target machine's kernel, must be preserved when it leaves the target's network on its way to the recursive resolver. This is not guaranteed, due to NAT/CGNAT devices on the target's network or its ISP network. A comprehensive research \cite{CGNAT} found that ``92\% of UPnP sessions are from devices behind port-preserving CPEs [Customer Premises Equipment].'' \cite[Fig. 8(b)]{CGNAT} and ``Port allocation strategies observed for CGN [Carrier Grade NAT] ASes'' has ``port preservation'' behavior in 41.2\% of the non Cellular networks and 27.9\% of the cellular networks \cite[Table 6]{CGNAT}.
We also note that for IPv6 networks, NAT is not needed (and hardly used), hence the issue of UDP/IPv6 port modification is probably very rare.

\subsection{Important Observations}
\subsubsection{The Systemd-Resolved DNS Stub Resolver}

\label{sec:systemd-resolved}
The Ubuntu Linux distribution uses Systemd-Resolved since Ubuntu 16.10. Systemd-Resolved delegates the choice of UDP source port for outbound DNS queries to the underlying operating system (Linux). From our experiments, Systemd-Resolved performs record-based caching. Also, while Systemd-Resolved matches the DNS answer UDP source port (implicitly), the DNS TXID and the DNS query part to the respective DNS query data, it {\em does not} perform any sanity checks on the answer part itself, and will blindly cache DNS records found there, even if they are completely unrelated to the query. 

When Systemd-Resolved is configured to use multiple DNS resolvers, e.g. {8.8.8.8} and {8.8.4.4}, the attacker cannot predict which name server Systemd-Resolved will use for the DNS query, because Systemd-Rersolved switches the ``current'' resolver when the latter responds with {\tt SERVFAIL} for two consecutive queries or when Systemd-Resolved times-out. When switching resolvers, Systemd-Resolved flushes its cache. Systemd-Resolved imposes a TTL cap of 2 hours.

\subsubsection{Recursive Resolver Timeout}
\label{timeout}
Of interest to us is the query timeout imposed by the recursive DNS resolver in use by the target of the attack. This timeout, if it is shorter than the stub resolver's timeout, can limit the number of spoofed DNS answers the attacker can send the target machine before the recursive resolver gives up and sends a {\tt SERVFAIL} answer (which closes the attack's window of opportunity for a particular query).

\subsubsection{The Exim (SMTP Server) Example}
We chose to experiment with Exim, as more than 57\% of SMTP server deployments are using it according to a recent survey \cite{smtp-survey}. The attack is by no means limited to Exim, and can likely be applied to any SMTP server, and in general, to any server that can be forced to emit DNS queries to an attacker controlled domain, with the necessary alterations. Some observations on Exim that are helpful in customizing the attack for Exim can be found in Appendix~\ref{app:exim}.

\section{Device Tracking (Linux and Android)}
\label{sec:device-tracking}
\subsection{Attack Setup}
The attack is based on a setup similar to the tracking attacks of \cite{KP18, KP19-usenix}: the device tracking is carried out across the Internet (web), using an HTML snippet which can be embedded in any web site. The snippet forces the browser to emit a burst of TCP/IPv6 (and in recent kernels -- also TCP/IPv4), UDP/IPv6 or UDP/IPv4 packets to a destination IP address under the attacker's control. Multiple bursts are needed to extract enough core-states in order to guarantee long term tracking. 

\subsection{Attack Description}
The attacker uses e.g. WebRTC STUN requests to send multiple UDP packets to the same (attacker controlled) destination IP address. Each STUN destination is assigned a different UDP destination port, thus enabling the attacker to observe the original order of transmission. The attacker collects the STUN requests, ordered by the destination ports, and uses the source ports to reconstruct the core-state of the PRNG for the core that generated the STUN UDP source ports. 
Due to core switching, which is fairly common, and due to the multitude of cores (in mobile devices, namely Android smartphones, up to 8 cores), the attacker needs to run multiple iterations of the above. In our proof of concept, we used 16 iterations, 1 second apart, thereby taking overall 16 seconds to complete the measurement. See Appendix~\ref{parameters-for-tracking} for a discussion on the choice of parameters.
The attacker extracts the core states of several PRNG instances -- with extracted states belonging to $\ge$5 different cores, it is possible to track an 8-core device. The attacker then carries out the observation linking algorithm described in Section~\ref{tracking} to associate the current device with an already seen device, or assign this device a new ID.

Of course, the above use of STUN UDP source ports is but an example (and our proof of concept implementation). It is possible to use the IPv6 flow label values from HTTP/HTTPS over TCP/IPv6 traffic, from gQUIC over UDP/IPv6 traffic (keep in mind that gQUIC in Google Chrome uses the UDP {\tt connect()} method, for which the Linux kernel employs the connection-oriented flow label generation algorithm), or from WebRTC TURN over TCP/IPv6 traffic. Other alternatives are to use UDP source ports using WebRTC TURN, and TCP/IPv4 ID e.g. using WebRTC TURN.

\subsection{Invariance to Choice of IP Address}
The technique does not make any assumption on the destination IP address. Hence, different HTML snippets can use different IP addresses, and moreover, even the same snippet can use multiple IP addresses and still get the same device ID for a given device. This makes it harder to detect the attack.

\section{Additional Attacks}
\label{sec:additional-attacks}
\textbf{Host Alias Detection \cite{10.1007/978-3-540-31966-5_9} and de-NATting \cite{Bellovin2002},\cite{Herzberg2017}:} By employing the tracking technique concept, we can establish which server IP addresses (including a mix of IPv4 and IPv6 addresses) belong to the same physical host, even if the server is behind a NAT, and even if the IP addresses are owned by different containers on the same host. We ran some successful tests over the Internet, including alias detection across containers, but in order to keep the paper short and focused, we omit the full discussion here.\\
\textbf{Local Attacks:} The attacks discussed above are all remote, but on Linux systems it may be possible to mount local attacks across users or across containers, which are more effective than remote attacks. 

\section{Experiments and Results}
\label{sec:experiments-and-results}
\subsection{DNS Recursive Resolver Timeout}
The impact of DNS recursive resolver timeout on the success probability of DNS cache poisoning is explained in Section~\ref{timeout}. In order to measure the timeout of popular DNS recursive resolvers, we conducted a simple experiment wherein we queried the recursive resolver for a host name in a domain we control, where the authoritative name server simply does not respond to the queries. The results are summarized in Table~\ref{tab:timeout}. The 5 selected public DNS services are the most popular ones according to \cite[Table 1]{public-dns-popular}.

\subsection{Measuring the UDP Port Offset of the DNS Query}
\label{sec:payload}

In preparation for the DNS cache poisoning attack (in Section~\ref{sec:poisoning}), we need to measure the offset of the DNS query source port generation point from the PRNG core-state we obtain. As an important by-product, we also measure the probability of the DNS query to be emitted from the same core as the one whose core state we extract, which provides us with an estimate of the DNS cache poisoning success probability.
The measurement setup is quite similar to the full fledged DNS cache poisoning experiment we conducted later (in Section~\ref{sec:poisoning}). It consists of a target system $\mathcal{T}$ (Ubuntu 20.04 Linux, with Linux kernel version 5.5.6-050506-generic, running on Dell Latitude E7450 computer with 4 logical cores Intel i7-5600U CPU at 2.6GHz), and a remote attack system $\mathcal{A}$. The target system uses the default Systemd resolver (version 245.4-4ubuntu3.1), with a recursive resolver $\mathcal{R}$ configured as Google Public DNS -- 8.8.8.8 and 8.8.4.4, and Exim version 4.93 as an SMTP server, in a default configuration. Other than our measurements, the target machine is idle. 
An attacker domain {\tt attacker.site} is also set up on an attacker-controlled authoritative name server $\mathcal{N}$ so that it does not respond to queries for DKIM TXT records (once per queried name; subsequent queries are answered immediately). 

Since our lab has no native IPv6 connection to the Internet, we encapsulated each IPv6 packet from the attacker system to the target system and vice versa with UDP/IPv4 packet (with random source port, so that each packet belongs to a different flow) in a proprietary encapsulation protocol between the attack system and an auxiliary host $\mathcal{X}$ on the target system's LAN. This host terminates the IPv6 encapsulation and sends and receives native IPv6 to/from the target system. Note that the encapsulated packets travel through the public Internet and are therefore subject to the same conditions the native IPv6 packets would have been subject to. For simplicity, we ignore this detail throughout the paper.

Additionally, in order to avoid sending spoofed DNS answers over the public Internet, the attacker system sends DNS answers with the genuine (attacker) source IP address, and we set up the target system to convert the source IP addresses locally to the name server's IP address using {\tt iptables SNAT} rules. 

We carried out 5000 measurements in the above system. A single measurement/iteration consists of the following steps (see also Fig.~\ref{fig:msc}):
\begin{enumerate}
    \item[1-3.] The attacker machine establishes a TCP connection (for the SMTP port) to the target machine.
    \item[4-7.] The attacker sends SMTP {\tt HELO}, {\tt MAIL FROM}, {\tt RCPT TO} and {\tt DATA} commands, with a few milliseconds delay between then, but without waiting for the target's response!
    \item[8.] The attacker sends a burst of 8 TCP SYN packets to the target machines, from different source ports, with a few millisecond delay between them, again, without waiting for the target's response.
    \item[9.] The attacker sends the SMTP message data (incl. DKIM header), followed by the End-of-Message marker.
    \item[10.] This triggers the target machine to emit a DNS query for the attacker's domain to the recursive resolver. The recursive resolver eventually forwards this query to the attacker's authoritative server (not shown in Fig.~\ref{fig:msc}).
    In the measurement experiment, we also capture this DNS query and record its UDP source port.
    \item[11.] The attacker waits for several milliseconds, then sends another TCP SYN packet (the ``post-DNS'' probe).
    \item[12.] The attacker now waits for the responses for all the packet he/she sent. The attacker analyzes the flow label values of the 8 SYN+ACK packets received earlier and extracts the core-state of the PRNG used to generate them (see Section~\ref{sec:cryptanalysis}). The last known core-state (right after the last flow label generation) is calculated. The attacker then calculates the PRNG offset (if it is found) of the flow label received for the post-DNS probe from the last known core-state.
    
    In the measurement experiment, we record the UDP source port of the DNS query, the PRNG offset (if found), and the post-DNS probe offset (if found).
\end{enumerate}

A single UDP port measurement can yield one of 3 results:
\begin{itemize}
    \item Failure to reconstruct a consistent core-state: this means that the flow labels were either generated non-sequentially (i.e. another {\tt prandom} consumer interfered with the sequence), or generated on multiple cores.
    \item No match: a failure to find a match to the UDP source port of the DNS query means that the DNS query UDP source port was probably generated on a different core.
    \item Match found: this is the ideal case, of course, and we record the offset in PRNG steps of the PRNG core-state used to generate the DNS query UDP source port, from the core-state at the end of the last flow label generation.
\end{itemize}
Our results for 5000 measurements are summarized in Table~\ref{tab:offset}. The  number of usable measurements is 4924 since the UDP port recording failed for 76 measurements. 
Denote by $O_U$ the offset of the PRNG when generating the UDP source port (from the PRNG state at the end of the pre-DNS probing), and denote by $O_P$ the offset of the PRNG when generating the flow label (or IPv4 ID, see below) for the post-DNS probe. 

As can be seen, guessing a PRNG step offset of 0 in the attack phase yields $p(O_U=0)=0.093$ success rate, given a correct guess of DNS TXID and server IP address.

We also measured $p(O_P \ge 4)=0.117$, $p(O_P \ge 4|P_U=0)=0.998$ and $p(O_U=0|O_P\ge 4)=0.796$.
This enables the improvement mentioned in Section~\ref{sec:dns-attack-basic}, namely we only attack the target system if $O_P\ge4$. Since $p(O_P \ge 4)=0.117$,
the saving in attack time and network resources is significant, and the cost (skipping successful attacks) is negligible. 

Using the data we obtained from the measurement runs, we can estimate the success parameters of the full attack experiment.
A single iteration runs steps 1-12, which take $t_\Delta+2 \cdot t_{\mathcal{A}\rightleftarrows\mathcal{T}}$ time, and can then either terminate, or mount an attack. From the attack script implementation, we have $t_\Delta=224ms$.\footnote{In the LAN case, we needed to introduce another delay, thus for LAN, $t_\Delta=314ms$.} An attack is mounted with probability $p(O_P \ge 4)$. Given the conditions for the attack are met, the attacker predicts the UDP source port with probability $p(O_U=0|O_P\ge 4)$. Assuming an attack window of $t_O+t_{\mathcal{T}\rightleftarrows\mathcal{R}}$, the actual attack burst can last $t_O+t_{\mathcal{T}\rightleftarrows\mathcal{R}}-t_{\mathcal{A}\rightleftarrows\mathcal{T}}$. Given bandwidth $b$ (in Mb/s), this allows sending $(t_O+t_{\mathcal{T}\rightleftarrows\mathcal{R}}-t_{\mathcal{A}\rightleftarrows\mathcal{T}})\cdot 10^6 \cdot b$ bits. Our spoofed DNS answers have 146 bytes (at the IP layer), so covering the entire TXID space requires $8 \cdot 146 \cdot 65536$ bits. Thus a single burst covers $0.01306 \cdot (t_O+t_{\mathcal{T}\rightleftarrows\mathcal{R}}-t_{\mathcal{A}\rightleftarrows\mathcal{T}}) \cdot b$ of the TXID space.\footnote{Given a fixed UDP source port, the upper limit of the number of packets the attacker can effectively send is $65536|\mathcal{R}|$, so the equations hold for $\frac{0.01306 \cdot (t_O+t_{\mathcal{T}\rightleftarrows\mathcal{R}}-t_{\mathcal{A}\rightleftarrows\mathcal{T}}) \cdot b}{|\mathcal{R}|} \le 1$. This condition is indeed met throughout our experiments.} 
The probability $p_B$ of the burst to poison the target's cache is, therefore: $$p_B=\frac{0.01306 \cdot p(O_U=0|O_P\ge 4)(t_O+t_{\mathcal{T}\rightleftarrows\mathcal{R}}-t_{\mathcal{A}\rightleftarrows\mathcal{T}}) \cdot b}{|\mathcal{R}|}$$
denote by $t_{\TOTAL}$ the total time needed until poisoning is achieved. Then, since the iterations are independent Bernoulli trials, we have:
$$E(t_{\TOTAL})=\frac{t_\Delta+2 \cdot t_{\mathcal{A}\rightleftarrows\mathcal{T}}}{p(O_P \ge 4) \cdot p_B} + \frac{t_O+t_{\mathcal{T}\rightleftarrows\mathcal{R}}-t_{\mathcal{A}\rightleftarrows\mathcal{T}}}{p_B}$$

For probes that use the TCP/IPv4 IPID field, a slightly different heuristic provides better UDP source port prediction. We found empirically that in this case, when looking only at $O_P\ge4$, $O_P-4$ is a better predictor for $O_U$. Indeed, $p(O_U=O_P-4)=0.021$,
whereas $p(O_U=0)=0.0081$.
We note that the results for TCP/IPv4 IPID are markedly inferior to those of TCP/IPv6 flow label. This due to the additional DNS queries incurred when the TCP handshake is completed (see Appendix~\ref{app:exim}). The additional queries, whose timing is somewhat non-deterministic, have two consequences: (1) they drastically reduce the probability of obtaining the correct linear equations, and hence the probability of obtaining a solution and thus a PRNG state drops to 0.261 (whereas it is 0.914 in the flow label case);
and (2) they result in much more dispersed values for $O_U$, $O_P$ and $O_P-O_U$. 

\begin{table}[]
\caption{DNS Query UDP Source Port Offset (4 Logical Cores)}
\label{tab:offset}
\begin{center}
\begin{tabular}{|l|r|r|r|}
\hline
Status & Offset & Count & Prob. \\ \hline
Failure & N/A & 424 & 8.6\% \\ \hline
No Match & N/A & 4006 & 81.4\% \\ \hline
\multirow{4}{*}{Match} & \textbf{0} & \textbf{457} & \textbf{9.3\%} \\ \cline{2-4} 
 & 1 & 23 & 0.47\% \\ \cline{2-4} 
 & 2 & 6 & 0.14\% \\ \cline{2-4} 
 & $\geq$ 3 & 8 & 0.19\% \\ \hline
\end{tabular}
\end{center}
\end{table}

\subsection{DNS Cache Poisoning}
\label{sec:poisoning}

\begin{table*}[t]
\caption{DNS Cache Poisoning Results (4 cores, 2 name servers)}
\label{tab:poisoning-results}
\begin{center}
\begin{tabular}{|c|c|c|c|c|c|c|c|c|c|c|}
\hline
\rule{0pt}{10pt}
\\[-1.5em]
Ex. & Loc. & Dist. [Km] & Exploited Field & IPv6 Traffic & $b$ [Mb/s] & $t_{\mathcal{A}\rightleftarrows\mathcal{T}}$ [ms] & \begin{tabular}[c]{@{}c@{}}$E(t_{\TOTAL})$\\(expected)\end{tabular}& \begin{tabular}[c]{@{}c@{}}$\overline{t_{\TOTAL}}$\\(measured)\end{tabular} & \begin{tabular}[c]{@{}c@{}}$E(t^\BRUTE_{\TOTAL})$ \\ (expected) \end{tabular} & \begin{tabular}[c]{@{}c@{}} $\overline{g}$ \\ (measured) \end{tabular} \\[1ex] \hline
1 & $A$ & $\approx 3300$ & IPv6 flow label & Encapsulated & 30 & 83.47 & 28.65s & 28.58s & 144114.3s & 4967 \\ \hline
2 & $A$ & $\approx 3300$ & IPv6 flow label & Encapsulated & 15 & 83.12 & 57.20s & 48.98s & 288228.7s & 5885 \\ \hline
3 & $B$ & $\approx 9600$ & IPv6 flow label & Encapsulated & 25 & 158.04 & 47.64s & 54.42s & 172937.2s & 3178 \\ \hline
4 & LAN & $0.001$ & IPv6 flow label & Native & 70 & 1.39 & 9.87s & 10.34s & 61763.3s & 5973 \\ \hline
5 & $A$ & $\approx 3300$ & TCP/IPv4 IPID & N/A & 30 & 83.08 & 127.06s & 114.66s & 144114.3s & 1257 \\ \hline
\end{tabular}
\end{center}
\end{table*}

The setup is similar to the one described in Section~\ref{sec:payload}, with the additional attack steps as following: 
\begin{itemize}
    \item[13.] Assuming the attacker can predict a UDP source port, the attacker sends, as rapidly as possible, as many spoofed DNS answers to the query as possible, with the IPv4 destination address set to the target IP address, the IPv4 source address set to one of the DNS recursive resolvers known to be used by the target system (e.g. pick one of 8.8.8.8 and 8.8.4.4 at random), the UDP destination port calculated from the PRNG core-state rolled forward to the most likely offset (0 -- see Table~\ref{tab:offset}), the UDP source port set to 53 (DNS), a running counter for the 16-bit DNS TXID field, 
    the DNS query section set to the DKIM TXT query ({\tt foo.\_domainkey.{\it name}.attacker.site}), and the DNS answer section containing two records: {\tt foo.\_domainkey.{\it name}.attacker.site CNAME success.attacker.site} and {\tt www.google.com A {\it attacker-IP}}. 
            After sending the burst of spoofed answers, the attacker can check whether there was any attempt to resolve {\tt success.attacker.site}. This indicates a successful poisoning, since the target system will only query for it if it accepts one of the spoofed DNS answers as valid. 
    \item[14.] The recursive resolver receives no answer from the attacker's authoritative server, and times out. It then sends a {\tt SERVFAIL} answer to the target, which arrives after the burst of the spoofed answers is received by the target.
\end{itemize}

We ran 5 DNS cache poisoning experiments. Each experiment involved running 5000 iterations, collecting success indications, attack's time-to-success, round trip times and effective bandwidth (which can be considerably lower than the endpoints' nominal bandwidth due to network bottlenecks in the path). Each experiment involved a different attack setup. Experiment 1 (our baseline) was conducted with $\mathcal{A}$ at Azure France (location $A$), across a continent from $\mathcal{T}$ (in our Israel lab). Experiment 2 is similar to Experiment 1, with the bandwidth throttled to 15Mb/s. In Experiment 3, we placed $\mathcal{A}$ at Azure US-East (location $B$), across the Atlantic from our lab. Experiment 4 had $\mathcal{A}$ on the same LAN with $\mathcal{T}$ (therefore, we could use native IPv6 traffic and forego $\mathcal{X}$) and finally,  Experiment 5 tested TCP/IPv4 IPID probing (instead of TCP/IPv6 flow label probing), thus we did not need $\mathcal{X}$ as well. 
The results (expected and average values) are summarized in Table~\ref{tab:poisoning-results}. This table also compares our results with brute force attacks in the same setup. For brute force attacks (assuming they need no delays), $E(t^\BRUTE_{\TOTAL})=(H-L+1)\frac{|\mathcal{R}|}{0.01306 \cdot b}$. The gain $g$ of using our DNS cache poisoning technique, compared with the brute force method is the ratio $\nicefrac{t^\BRUTE_{\TOTAL}}{t_{\TOTAL}}$. Table~\ref{tab:poisoning-results} also depicts $E(t^\BRUTE_{\TOTAL})$, and the average gain factor $\overline{g}$ based on the actual poisoning timing.

Denote the number of spoofed packets sent by by $M$. Then we have $E(M)=\frac{65536|\mathcal{R}|}{p(O_U=0|O_P \ge 4)}$, and  $E(M^\BRUTE)=65536(H-L+1)|\mathcal{R}|=3700424704$ packets. In e.g. Experiment 1 (transcontinental), $E(M)=164629$ and we measured $\overline{M}=166892$ packets -- almost 4.5 orders of magnitude less than a brute force attack. The average number of iterations in this case was 56.1.

\subsection{Device Tracking via UDP/IPv4 Source Port}
\label{sec:tracking-experiment}
We set up a demo tracking site in Azure, and used it to measure devices and networks. The demo site is built around an HTML snippet and Javascript code as explained in Section~\ref{tracking}. Particularly, we used $N=1{,}000{,}000$ (this limited the longevity of tracking, but allowed us to take some coding short-cuts), 16 measurement iterations, 1 second apart, and in each iteration, we had the browser send 31 WebRTC STUN requests to consecutive destination UDP ports on a single destination IPv4 address on our server. The browser had to dwell on our web site for about 16 seconds for the measurements to conclude. At the server side, we extracted the core-states of the device by assuming the Google Chrome PRNG usage pattern (every other PRNG invocation) and extracting the core-state from the sequence of consecutive UDP source ports in each iteration, assuming no interference and no core-switching. Of course, not all iterations yield a valid core-state, mostly due to core-switching, but typically, enough iterations yield a valid core-state, and with enough core-states, it is guaranteed that a match will be found to a previous core-state extracted earlier from the same device.
In case no core-state is extracted in all 16 iterations, it is quite likely that there is a network-level interference. We found that in all such cases, the UDP source port was overridden by an intermediate network device. This was easy to verify, as we know the 3 port ranges used by Linux/Android, and assuming no interference, the source ports should be uniformly distributed in one of the ranges. However, in all cases where no core-states were extracted, some source ports were either outside all three ranges, or the source ports were distributed in a very non-uniform manner (e.g. clustered in small sub-ranges) that indicates UDP source port overriding. The single exception is one network that completely dropped our WebRTC STUN packets.

Our experiment involved having the owners of 13 devices (Android mobile phones from Google, HTC, Huawei, OnePlus, Pocophone and Samsung, and Linux Dell laptops) browse to our demo website from various networks. For 11 devices, the core-state extraction worked when the network did not interfere. Two devices were only measured through interfering networks, and thus they are not counter examples. 

We surveyed 41 IPv4 networks (airport/hotel/conference/restaurant WiFi networks, cellular networks, enterprise WiFi networks, etc.) across multiple countries (Canada, Germany, Israel, Italy, United Kingdom and USA). In 24 networks out of 41 (59\%), our core-state extraction technique succeeded, which indicates that the UDP source ports were preserved.

Of the 36 landline networks we tested, 23 preserved the UDP source ports -- among them AT\&T (US), Azure (US), BezeqInt (IL), FastWeb (IT), LeaseWeb (DE), Level3 (US), SoftLayer (US), Telus (CA)  and Vodafone (DE). 13 networks did modify the UDP source ports -- among them Bell Canada (CA) and Shaw Communications (CA). Of the 5 cellular networks, only one network (in roaming mode) preserved the UDP source ports, while the remaining 4 networks modified the UDP source port, among them Hutchison (UK).

We can thus say that the majority of the IPv4 networks we surveyed (59\%) do not modify the UDP source port, and are therefore conductive to device tracking via UDP source ports, and to DNS cache poisoning via UDP source port prediction. We did not test IPv6 networks, but it is likely that UDP source port modification is rare in IPv6 since NAT is not usually needed for IPv6 networks. 

We were even able to measure and identify devices that used ExpressVPN and TunnelBear VPN solutions. Our device tracking technique worked across networks and VPNs, i.e. we identified the same device as it emerged via different TunnelBear exit nodes, and also without TunnelBear VPN, and likewise with ExpressVPN.

We also verified that browsing with Google Chrome's ``incognito mode'' does not affect our tracking technique, i.e. the device is still tracked and still associated with the previously measured core-states, whether they were measured in ``incognito mode'' or not.

Appendix~\ref{app:additonal-device-tracking-results} contains additional experiments and results.

\section{Discussion}
\subsection{DNS Cache Poisoning}
\label{sec:dns-poisoning-discussion}
The DNS cache poisoning experiments demonstrate that our technique reduces the average poisoning time from an order of days to less than a minute. Arguably, even without our attack, using brute force is still a viable threat. However, our results are an {\em enabler} to the following:
\begin{itemize}
    \item The time scale of a brute force attack (days) allows a defender to incorporate a human in the loop, for example, to analyze and to manually adjust thresholds and drop lists. The time scale of our attack (less than a minute), defeats human intervention.
    \item The brute force attack packet count (billions) allows a defender to set a very high and coarse grained threshold for e.g. spoofed/unexpected packets. Our attack sends only 160,000-170,000 packets on average, and may ``fly under the radar'' for these thresholds.
    \item Our attack allows the attacker to poison using multiple payloads (packets) in a short time frame (66-124 packets/hour). A brute force attack can only poison using a single payload at a time (due to cache expiration). So for example, if the attacker needs to poison several targets almost simultaneously (within minutes), this is possible using our technique, but practically impossible using brute force.
\end{itemize}

Note: our experiments exploit the predictability of the UDP source port, when it is not overridden en-route. When the UDP source port is modified in transit, this may open another venue for DNS cache poisoning attacks, since the modified source port may either be predictable in itself, or picked from a small pool of values (instead of roughly 30,000 values). This is hinted to by \cite{CERT-DNS}. We comment (without further investigation) that we saw many cases wherein UDP source ports were either sequential or selected from a small pool of values. 

\subsection{Workarounds for the DNS Cache Poisoning Attack}
\label{sec:workarounds}
By default, the DNS traffic between the stub resolver and the recursive resolver is not protected, i.e. the DNS traffic is sent in plaintext, with no integrity protection other than the UDP and DNS supposedly unpredictable parameters. Recently, several proposals were made to close this security gap, by encrypting and/or cryptographically validating the DNS data in transit between the stub resolver and the recursive resolver.
\subsubsection{Using a (DNSSEC) Validating Stub Resolver}
DNSSEC \cite{RFC4033} can be used to protect the integrity of the DNS records queried by the stub resolver, if the stub resolver supports DNSSEC validation (and of course, as long as the entire DNS hierarchy for the queried record at hand supports it).
Systemd-Resolved supports DNSSEC validation via a flag. 
glibc stub resolver (both the in-process resolver and the NSCD daemon) does not support DNSSEC validation.
\subsubsection{DNS over TLS (DoT) and DNS over HTTPS (DoH)}
DoT \cite{RFC7858} and DoH \cite{RFC8484} both employ end-to-end encryption and integrity checking between the stub resolver and the recursive resolver, using TLS and HTTPS, respectively. This requires the recursive resolver to support DoT/DoH.
Systemd-Resolved supports DoT by setting via the DNSoverTLS flag. 
glibc apparently does not support DoT and DoH. 
\subsubsection{DNS Transaction Signature (TSIG)}
TSIG \cite{RFC2845} is DNS-related protocol that provides cryptographic signatures to DNS answers. Systemd-Resolved and glibc do not support TSIG.
\subsubsection{0x20 Bit-Encoding Queries} 
Using 0x20 Bit-Encoding Queries \cite{0x20} by the stub resolver can increase the entropy of queries, thereby increasing the attack's time, ideally to such magnitude that makes the attack impractical. This is effective for attacks against query-based stub resolvers, but not against record-based stub resolvers (in their case, the attacker's authoritative name server observes the additional entropy bits and renders the attack almost ineffective). 
\subsubsection{Changing Stub Resolver}
Switching to a stub resolver which chooses the UDP source port of outbound DNS queries using a cryptographically secure random number generator (rather than delegating this choice to the underlying operating system -- Linux in this case) eliminates the DNS query UDP source port prediction attack we described against Linux.

\subsection{Workaround for Device Tracking}
By using a forward proxy server, the browser delegates network connection establishment to the proxy server, thereby not exposing the device's network protocol fields to the target website. This solution only works if UDP traffic is forwarded through the proxy 
or if the browser's UDP traffic is dropped.

\subsection{Recommendations}
The root cause of the problems we described above is the use of a non-cryptographic PRNG to generate visible protocol fields -- even when the field itself carries no security significance (like IPv6 flow label) this can still be abused for device tracking; and when the field does have security significance (like UDP source port), the use of non-cryptograhic PRNG is an obvious flaw. The issue is exacerbated by the partial (only) re-seeding and the sharing of the PRNG instance among multiple consumers (including across containers). 
Our recommendations are, therefore, to use an industrial-strength PRNG with a sufficiently random key/seed, and to ensure that PRNG invocation and re-seeding is interrupt- and thread-safe.

\subsection{Vendor Status}
\label{sec:vendor-status}
We notified the Linux kernel team on March \nth{8}, 2020. A ``band-aid'' fix (commit f227e3ec3b5c) was deployed to versions 5.8, 5.7.14, 5.4.57, 4.19.138, 4.14.193, 4.9.233 and 4.4.233. A full patch (commit c51f8f88d705) was deployed to versions 5.10-rc1, 5.9.2, 5.4.78, 4.19.158, 4.14.207, 4.9.244 and 4.4.244.
This issue is tracked as CVE-2020-16166.

\section{Conclusion}
We analyzed the Linux kernel {\tt prandom} PRNG and found several weaknesses in it -- the PRNG is predictable since it is completely linear; it is only partially re-seeded; and the same set of instances are used for all PRNG ``consumers'', including across containers. As there are PRNG consumers in multiple network layers, it is possible to extract data from one protocol and use it to exploit another protocol in a different layer, hence demonstrating a ``cross layer attack''. Our main result is a DNS cache poisoning attack against Linux, which is a cross layer attack (from IPv4/IPv6 to UDP). Additionally we describe a device tracking attack against Linux and Android.

Our attacks are practical. In our experiments, the DNS cache poisoning attack takes on average 28.6 seconds (trans-continental attacker) and 54.4 seconds (transatlantic attacker) with a modest Internet connection. The device tracking technique takes 16 seconds (this can probably be reduced even further), and is very scalable.

While the attacks were demonstrated using UDP/IPv4 where source port preservation is not guaranteed, it is possible to mount the same attacks over UDP/IPv6 where UDP source port overriding is highly unlikely, thus achieving coverage of almost 100\% of the networks. Device tracking and network intelligence attacks can be also be carried out over TCP/IPv4 (for recent kernels) using the IP ID field which is unmodified in 92\% of the networks.

\section*{Acknowledgment}
We thank Jonathan Berger, Benny Pinkas and the anonymous S\&P 2021 reviewers for their helpful suggestions and comments.
This work was supported by the BIU Center for Research in Applied Cryptography and Cyber Security in conjunction with the Israel National Cyber Bureau in the Prime Minister's Office, and by the Alter Family Foundation.

\bibliographystyle{IEEEtran}

\appendices
\section{Details of the Baby-Step-Giant-Step Application}
\label{app:baby-step-giant-step}
Assuming we have a readout of $S_2, S_3, S_4$ from an examined device. We use the baby-step giant-step technique \cite{1971-shanks} to find a previous observation of the device, if such exists. For this, we maintain a table $T$ in memory that maps the next $\sqrt{N}$ giant steps (each such step consists of $\sqrt{N}$ baby-steps) from previous observations of up to $R$ already observed devices, to their device IDs. We then calculate the next $\sqrt{N}$ (baby) steps of $(S_2,S_3,S_4)$, including the initial state itself -- each such vector $V_j$ has $k_2+k_3+k_4=82$ independent bits. We look up each $V_j$ in $T$. If a match for $V_j$ is found, then the device ID returned is $T[V_j]$. If no match is found for any $V_j$, then this is deemed a new device, and we generate a new (unique) device ID for it -- $\ID$. This can be done e.g. using a counter, a fine-grained timestamp, or a PRNG. $T$ is then updated in the following manner: we advance $S_2, S_3, S_4$ $\sqrt{N}$ times (this giant-step can be done by multiplying by a matrix representing a single-step matrix raised to the power of $\sqrt{N}$, or even further optimized by keeping a map in memory from all possible $2^{k_2}$ $S_2$ values to their values after $\sqrt{N}$ steps, and same for $S_3$ and $S_4$) to form $V^{\ID}_0$, and insert $(V^{\ID}_0 \rightarrow \ID)$ to $T$. We do this $\sqrt{N}$ times to form $V^{\ID}_j, j=0,\ldots,\sqrt{N}-1$, inserting key/value pairs $(V^{\ID}_j \rightarrow \ID)$ to $T$ correspondingly. Finally, $\ID$ is returned as the device ID.

The lookup time is therefore $\sqrt{N}$, with false positive probability bounded from above by $\frac{RN}{2^{k_2+k_3+k_4}}$. The hash table insertion time is $\sqrt{N}$. The memory size $|T|$ needed is $R\sqrt{N}$ entries. 

The algorithm correctly identifies an already-seen device: if the device was already seen before, i.e. has device identifier $ID$, and the current state is $0<n<N$ steps after the state it was seen before then we can write $n=a\sqrt{N}+b$ (with $0 \leq a,b < \sqrt{N}$), and it is easy to see that $V^{ID}_a=V_{(\sqrt{N}-b) \mod \sqrt{N}}$.

In the multi-core device scenario, assume a device has $C$ cores, and assume that during every sampling session, through repeated application of the algorithm described in Section~\ref{sec:cryptanalysis}, we obtain core-states from at least $c=\lfloor \frac{C}{2} \rfloor +1$ different cores. Since $c>\frac{C}{2}$, we are guaranteed to have at least one {\em common} core -- a core whose core-state is extracted in the present measurement and in the previous observation of the same device, if a previous observation exists. Therefore, we simply need to run the algorithm above for each core-state extracted from the device. This has the impact of multiplying $R$ by $C$ (as well as the run-time for each device). Since all the devices we encountered have $C \leq 8$, this is not a very significant factor. The above discussion assumes a uniform distribution of cores used during the PRNG consumption. This is a worst case scenario for the attacker, since a non-uniform distribution needs fewer iterations to statistically guarantee common core between two observations.

Two random, unrelated devices can get ``accidentally'' linked with probability of $\frac{2N-1}{2^{k_2+k_3+k_4}}$, therefore the device space entropy is $\log_2\frac{2^{k_2+k_3+k_4}}{2N-1}=k_2+k_3+k_4-1-\log_2N$ bits (as $\log_2{2N-1} \approx 1+\log_2N$ when $N \gg 1$), and therefore, when $\log_2\binom{R}{2} \ll k_2+k_3+k_4-\log_2{N}-1$, the probability of finding {\em any} collision is low.

\section{Relative ``Drift'' in LFSR States of the Same Core}
\label{app:drift}
Another consequence of the non interrupt-safe implementation of the {\tt prandom} PRNG is that it can happen that an interrupt handler that invokes the PRNG is executed due to an interrupt in such way that de-synchronizes the $S$ states. In normal circumstances, each $S_i$ in the same core-state advances once per PRNG invocation. However, suppose that a thread enters the PRNG code, reads $S_1$ and $S_2$ atomically, as a single 64 bit quantity, then an interrupt is triggered, and the interrupt handler invokes the PRNG code which reads $S_1, S_2, S_3, S_4$ and updates them. Finally, once the interrupt handler finishes running, the kernel switches back to the thread which reads $S_3, S_4$ atomically and updates $S_1, S_2, S_3$ and $S_4$. The PRNG core-state after this execution sequence will be such that $S_1, S_2$ are advanced once because the thread updates them based on its registers before the interrupt, but $S_3, S_4$ are advanced twice -- first by the interrupt handler, then by the thread. This phenomenon causes a relative ``drift'' between the LFSRs in the same PRNG core-state. 

This relative ``drift'' plays a major role in the device tracking scenario. 
When a relative drift is possible, one can no longer assume that all $S_2,S_3,S_4$ advance together, and therefore the algorithm presented in Section~\ref{tracking} cannot be applied as-is. We only encountered such drift on a Samsung A50 mobile phone, so this phenomenon is probably model-specific, and as such, an attacker can apply the drift-aware algorithm only for models that require it (e.g. using the User-Agent header). 

We now describe a variant of the device identification algorithm that can handle the drift scenario. We assume that with up to $N$ PRNG steps, the drift is up to $K$ PRNG steps. The variation is that the table $T$ contains the states for $S_3$ and $S_4$ only, mapping to the tuple (or tuples) $(ID,S_2)$. The initial lookup in $T$ is identical to the original algorithm, and results in $\frac{RN}{2^{k_3+k_4}}$ ``random'' matches. Each match also offers the progress $n$ of $S_3,S_4$. We then take the $S_2$ value from $T$, advance it $n-K$ steps, and then advance it $2K$ times more, with each step comparing the value to the $S_2$ extracted from the machine. We expect a random match probability of $\frac{2K}{2^{k_2}}$ in this step, and overall expected $\frac{2K \cdot R \cdot N}{2^{k_2+k_3+k_4}}$ random matches. If a match is found, we deem that the device has identifier $ID$. If not, we add the mappings $(S_3,S_4) \rightarrow (ID,S_2)$ to $T$ for all $\sqrt{N}$ giant steps of $S_3,S_4$ (but without changing $S_2$).

From our measurements, typically $K \approx \frac{N}{5000}$, so with the above assumptions ($N=10^7$) we have $K=2000$. This algorithm works well if $K \cdot R \cdot N \ll 2^{k_2+k_3+k_4}$, which is the case for $K=2000$. In terms of run time, it only adds $2K$ operations per lookup, And in terms of memory, it is comparable to the original algorithm.

The entropy for the drift scenario has $\log_2(2K)$ less bits than the original algorithm, i.e. with $R$, $N$ as in Appendix~\ref{app:baby-step-giant-step}, the entropy becomes 44.781 bits.

\section{Attacking Query-Based Caching Stub Resolvers}
\label{app:attacking-query-based}
The DNS cache poisoning attack described in Section~\ref{sec:dns-attack-basic} is still applicable (with some necessary changes) to query-based caching DNS resolvers, such as glibc's NSCD caching DNS stub resolver, though it is not as efficient as the attack against record-based caching DNS stub resolvers. For query-based caching resolvers, the attacker should force the server to emit a DNS query for the host name which is the target of the poisoning attack, e.g. if the attacker wants to poison the cached DNS resolution of www.google.com in the target system, then the attacker needs to force the target machine to emit a DNS query for www.google.com. For this, the attacker cannot use the DKIM technique, since the queried name will have {\tt \_domainkey} as one of its labels. However, we noted during the experiment described in Appendix~\ref{app:public-dns-use} that some MTAs do send MX queries for the sender domain, and these queries can thus be used for DNS cache poisoning. Of course, cache poisoning only succeeds if the DNS resolution for www.google.com is not already cached (or is stale) by the stub resolver. And since a failure to poison the cache for the query at hand causes the genuine resolution to be cached, the attack can only take place once every TTL seconds (where TTL is the DNS ``time to live'' value for a genuine answer to the query at hand). As such, the attacker should ideally target names whose genuine TTL is relatively low. For example, the TTL for www.google.com's A record is 300 seconds. Unlike the attack against record-based caching stub resolvers, here the attacker cannot increase the window of opportunity for his/her spoofed packets to win the race against the genuine recursive resolver answer. Therefore, the attacker is constrained by $t_{\mathcal{T}\rightleftarrows\mathcal{R}}$, with the possible addition of $t_{\mathcal{R}\rightleftarrows\mathcal{G}}$ in case the name is not cached at the recursive resolver. 
For the original attack to be effective, we require $t_{\mathcal{A}\rightleftarrows\mathcal{T}} \ll t_{\mathcal{T}\rightleftarrows\mathcal{R}}+t_{\mathcal{R}\rightleftarrows\mathcal{G}}$ or $t_{\mathcal{A}\rightleftarrows\mathcal{T}} \ll t_{\mathcal{T}\rightleftarrows\mathcal{R}}$ (for records cached in the recursive resolver). However, this is difficult to achieve in practice. Therefore, we must forego the ``post DNS'' probe improvement, and reorder the attack steps so that the pre-DNS probes return their results (and the attacker can calculate the PRNG state and predict the UDP source port) before forcing the target to emit the DNS query. This can be done in parallel to the ``main'' SMTP session, in order to save time. The final steps of the attack are then to force the target to emit a DNS query, and immediately send a multitude of spoofed DNS answers using the UDP source port predicted from the PRNG state -- as many answers as can be fitted into $t_{\mathcal{T}\rightleftarrows\mathcal{R}}$ or $t_{\mathcal{T}\rightleftarrows\mathcal{R}}+t_{\mathcal{R}\rightleftarrows\mathcal{G}}$ time frame. Unlike the original scheme, in this variant, every iteration ends in an attack (unless the PRNG state cannot be extracted) since there is no ``post DNS'' probe that can indicate that the process switched to a different CPU core.

\section{Attacking Cache-Less Clients}
\label{app:attacking-cacheless}
CentOS Linux and Debian Linux do not employ DNS cache, but it is still possible to attack them. There are scenarios in which spoofing a single DNS answer to a process can compromise the security of the target system. For example, consider a target system that runs a caching HTTP forward proxy service for an organization. An attacker can force a browser in the organization to render an attacker web page which forces the browser to request a page with a short life span, such as a news page. We will assume this page URL is http://www.google.com/news. Assuming this URL is not cached by the proxy service, or has become stale, the proxy service will try to retrieve the URL, by first resolving www.google.com. Assuming the attacker succeeds in spoofing the DNS answer, the forward proxy requests the said URL from the attacker's IP address and {\em caches} the HTTP response it gets from the attacker. Therefore, the attacker attains a long-lasting HTTP cache poisoning impact on the forward proxy -- all the proxy clients will be served the poisoned page in response to a request for http://www.google.com/news.

The DNS poisoning attack against a cache-less stub resolver is identical to the attack against query-based caching DNS stub resolvers in Appendix~\ref{app:attacking-query-based}, except that the impact is not on the target system DNS cache, but rather on a single DNS resolution requested by a single process. Additionally, there is no need to pick a low TTL host name to poison, and there is no need to wait TTL seconds between attempts, since the stub resolver has no cache to consult, and will thus {\em always} forward the query to the recursive DNS server. Therefore, this attack can run almost continuously. The attacker can interact with any server software/service (not necessarily the service whose resolution is targeted) to obtain a PRNG core-state. This takes $t_{\mathcal{A}\rightleftarrows\mathcal{T}}+t_\Delta$ time. Then the attacker forces the target service to emit a DNS query to the desired host name. Finally, the attacker can send a $t_{\mathcal{T}\rightleftarrows\mathcal{R}}$ burst of spoofed answers  -- as many spoofed packets as possible given the Internet bandwidth at the attacker's disposal or the target's Internet connection, and reiterate indefinitely. In each such iteration, the probability to successfully predict the UDP source port is $\approx \frac{1}{11}$, and the iteration time is $t_{\mathcal{A}\rightleftarrows\mathcal{T}}+t_\Delta+t_{\mathcal{T}\rightleftarrows\mathcal{R}}$, whereas in a brute force attack, each iteration has probability to correctly guess the UDP source port of $\frac{1}{H-L+1}$ and takes $t_{\mathcal{T}\rightleftarrows\mathcal{R}}$ time. Thus the gain is:
$$g=\frac{H-L+1}{11} \cdot \frac{t_{\mathcal{T}\rightleftarrows\mathcal{R}}}{t_{\mathcal{A}\rightleftarrows\mathcal{T}}+t_\Delta+t_{\mathcal{T}\rightleftarrows\mathcal{R}}}$$
For example, using the values measured in Experiment 1 (Table~\ref{tab:poisoning-results}), and using the average Google DNS response time for A records over IPv4 of 55.29ms \cite[Table 1]{aldalky2020revisiting}   ($t_{\mathcal{T}\rightleftarrows\mathcal{R}}=55.29ms, t_{\mathcal{A}\rightleftarrows\mathcal{T}}=83ms, t_\Delta=224ms$), we get $g=391.69$, which is 2.5 orders of magnitude faster than a brute force attack.
Note, however, that unlike a brute force attack, our attack cannot benefit from having multiple in-flight DNS queries, since our attack attempts to predict the UDP source port of the (first) query, and thus subsequent in-flight queries (whose UDP source port is even more difficult to predict) are mostly useless for our attack.

\section{DNS Cache Poisoning Impact}
\label{app:dns-cache-poisoning-impact}
Here are some examples for the possible DNS cache poisoning impact on various applications/services, assuming DoH/DoT is not used by these applications (see Section~\ref{sec:workarounds} for a discussion of DoH/DoT):
\begin{itemize}
    \item Hijacking emails: in case MTA-STS is not employed by the SMTP MTA, the attacker can force the MTA to interact with the attacker's endpoint, and hijack outbound emails. 
    \item MTA STS suppression: MTA STS relies on retrieving a TXT DNS record {\tt \_mta-sts.{\it domain}} in order to activate STS for {\it domain}. Such record can be poisoned to contain an invalid value which suppresses STS for the domain.    
    \item Hijacking browser (HTTP) traffic.
    \item Poisoning HTTP proxy cache.
    \item Poisoning DNS-based security records -- DNSBL, SPF, DKIM and DMARC. This, in turn, may enable email sender spoofing, which can facilitate more successful spam, phishing, email fraud and malware campaigns.
    \item Poisoning reverse DNS (PTR records) answers -- can manipulate logs, and in case there is authentication/authorization logic that depends on reverse DNS -- it can be bypassed.
    \item Local DoS (blackhole update services, important/sensitive hosts). For example: preventing access to Gmail servers and to security.ubuntu.com.
    \item NTP time-shifting attacks: as described in \cite{NTP-attack}, an NTP client relies on DNS to discover NTP servers, and therefore DNS cache poisoning can direct the NTP client to attacker controlled NTP servers.
\end{itemize}

\section{Detecting and Measuring the Use of Public DNS}
\label{app:public-dns-use}
\subsection{Detecting the Use of a Public DNS Service}
An authoritative name server can detect that the recursive resolver interacting with it is a public DNS service, through inspecting the source IP address of the DNS query. In our experiments from multiple locations (including our lab, location $A$ and location $B$), we observed that for the 5 most popular services (Google DNS, Cloudflare, IBM Quad9, Yandex.DNS  and OpenDNS), the source IP address is in an AS which belongs to the respective service (except for IBM, which uses WoodyNet AS). Thus detecting a public DNS is straight-forward -- obtain the AS name of the query's source IP address, un-capitalize letters, and search for the strings ``google'', ``cloudflare'', ``woodynet'', ``yandex'' and ``opendns''. A match indicates, with high likelihood, that the recursive resolver is a public DNS service.

\subsection{Measuring the Use of Public DNS Services}
We conducted an experiment based on the Alexa Top 1M domains list. We sampled 1000 domains in various offsets in the list. For each domain in a block of 1000 domains, we first extracted the highest priority MX record for the domain, filtered its domain against known \nth{3} party mail providers (e.g. google.com and googlemail.com for Gmail), then we made sure that the MX host name is ``related'' to the domain name (the domain SLD\footnote{SLD is ``Second Level Domain'', which is the distinctive domain name, e.g. the string ``google'' in \textbf{google}.com.} is a sub-string of the MX host name SLD) -- this indicates an {\em organic} host (a host that belongs to the domain's organization, as opposed to a \nth{3} party host). We also verified that the MX host IP is {\em not} in an AS that belongs to one of the top 5 public DNS services, and removed duplicate IP addresses. Then we established an SMTP session with the MX host, attempting to send a single message from test@{\it our-domain} (where {\it our-domain} is unique per-test sub-domain of a domain we own) to postmaster@{\it domain} with a DKIM header pointing at a record in {\it our-domain}. We recorded DNS queries for {\it our-domain} which is served by our authoritative name server. 

The above filtering procedure may incur false negatives (leaving out some organic servers), but the amount of false positives appears to be negligible. We note that some servers did not send queries to our domain, this is due mostly to anti-spam IP-based measures. We also note that some servers sent DNS queries even before the DKIM header was sent to them, e.g. trying to find more information about the sender domain, such as its MX record.

\begin{table}[]
\caption{Public DNS Service Statistics (Alexa Top 1K Domains)}
\label{tab:public-dns-experiment}
\begin{center}
\begin{tabular}{|l|r|}
\hline
Status & Count \\ \hline
No/invalid MX record & 176 \\ \hline
\nth{3} party mail service & 534 \\ \hline
SLD mismatch & 111 \\ \hline
TCP connection failed & 12 \\ \hline
Duplicate IP address & 5 \\ \hline
\nth{3} party MX IP address & 0 \\ \hline
No queries & 78 \\ \hline
Query from non-public DNS & \textbf{64} \\ \hline
Query from public DNS & \textbf{20} \\ \hline
\end{tabular}
\end{center}
\end{table}

Table~\ref{tab:public-dns-experiment} shows detailed break-down of our results for domains 1-1000 (the top 1000 domains). Of the 84 hosts which sent DNS queries, 20 (24\%) were from public DNS services. For domains 1001-2000, 37\% of the queries were from public DNS services, for domains 10001-11000 -- 36\%, for domains 100001-101000 -- 34\% and for domains 500001-501000 -- 35\%.

We conclude that public DNS services are used by 24\%-37\% of the {\em organic} hosts, with lower probability matching more popular (larger) domains.

\section{Observations on the Exim SMTP Server}
\label{app:exim}
We observed that upon a successful completion of a TCP handshake with a client, Exim emits a ``reverse DNS'' query (querying for a DNS PTR record) for the client network address. This DNS query consumes {\tt prandom} values, and therefore interferes with our PRNG core-state extraction method which relies on obtaining consecutive values of {\tt prandom} via IPv6 flow label values. Therefore, when the attacker sends a burst of TCP SYN packets to Exim in order to collect flow labels and extract a core-state of the PRNG, the attacker has to bypass the standard operating system APIs, craft his/her own TCP SYN packets and send them directly, so that there will {\em not} be a handshake conclusion. That is, the attacker sends TCP SYN, the server responds with TCP SYN+ACK, but the attacker never sends an ACK back. This way, Exim does not send the reverse DNS query. This, however, leads to TCP SYN+ACK packet re-transmissions from the target system, due to the incomplete TCP handshake state in which it ends up. These transmissions can interfere with the core-state extraction procedure, since each re-transmission generates a new flow label value, and therefore consumes another {\tt prandom} value. Hence, immediately upon receiving a TCP SYN+ACK from the target system, the attacker needs to send a TCP RST for the circuit, thus tearing it down before re-transmissions occur. 

We note that when specifying an attacker domain in a DKIM email header, Exim by default queries for the domain's DKIM public key (a DNS TXT record). The DNS query is sent roughly 8 milliseconds after the end-of-message (EOM) notification arrives from the client (attacker), in the form of data line containing a single dot. 

Finally, it is recommended to keep the SMTP connection ``frozen'' at the EOM state (and not conclude it with {\tt QUIT} or terminate the underlying TCP connection) until the DNS query is emitted. Terminating the SMTP connection typically results in immediate consumption of additional {\tt prandom} values, which, as explained above, is detrimental to the success of the attack.

\section{Choice of Parameters for Device Tracking}
\label{parameters-for-tracking}
The device tracking technique has several parameters:
\begin{itemize}
    \item Denote by $I$ the number of iterations.
    \item Denote by $L$ the number of UDP packets per iteration.
    \item Denote by $t_W$ the wait time between iterations.
\end{itemize}
Due to a bug in Chrome \cite{chromium-webrtc-bug}, the total number of WebRTC invocations in a short period of time is limited to 500. Therefore $L \cdot I \leq 500$. Thus we set $I=\lfloor \frac{500}{L} \rfloor$.

Experimentally, we obtained the range of effective $k$ values that are useful for core state extraction from UDP source port sequences. This range is $8 \leq k \leq 13$. Ideally, we would choose $L=29$, since for $k \geq 8$ it guarantees that if there is a single core-switch in the sequence, then at least one sub-sequence is generated by the same core -- for this we need $k \cdot \lfloor \frac{L+1}{2} \rfloor \geq 113$. However, due to some technical constraints in our linear algebra routines, we had to increase it a bit and use $L=31$. This sets $I=16$.

We set $t_W$ to a minimum value which takes into consideration the round-trip time between the measured device and the tracking website, and enough time (about 0.5s) for the tracking page to load into the browser, get parsed, execute the Javascript and send the STUN/TURN packets. We found empirically that $t_W=1s$ works, though this can probably be lowered a bit by fine-tuning the system.

\section{Additional Device Tracking Results}
\label{app:additonal-device-tracking-results}
\subsection{PRNG Consumption Rate}
\label{app:consumption-rate}
The demo setup of our tracking technique enabled us to measure the core state progression of the PRNG in the cores we extract. This is a by-product of the tracking technique, since for tracking, we need to associate the presently extracted core-state with a core-state extracted in the past, and the tracking technique does so by ensuring that the present core-state is $n<N$ steps further than the previously observed core-state. This $n$ is the PRNG progress since the previous core-state measurement.

We collected data on the core state progression on two regularly used Android devices. On a Samsung Galaxy A50 device,
we measured the PRNG core states of 5 distinct cores. The average progress measured was 15935.68-21343.01 steps per day over a period of around 5 days, depending on the core (core usage is not uniform). 
On a Samsung Galaxy Note 8 device, we measured average progress of 52292.81 steps per day over a period of 30 hours on a single core.
On a Google Pixel2 XL lab device which is not regularly used, we measured average progress of 14605.03-31699.91 steps/day over a period of almost 2 months, on 8 cores. 

Therefore, we can empirically bound from above the number of steps/day for a single core at 100,000.

\subsection{TCP/IPv4 IPID Device Tracking}
\label{app:TCPIP-IPID-tracking}
Using the same UDP source port device tracking framework, we set up a TCP/IPv4 IPID device tracking demo, using WebRTC TURN URL to establish a burst of TCP/IPv4 connections to the tracking web server, and collecting IP ID values from the TCP SYN packet. The PRNG state is reconstructed from the 16-bit IP ID values (similar to the way it is reconstructed from UDP source ports), and the remaining tracking functionality is identical to the UDP source port case.

Since IPID-based tracking only applies to recent kernels, we could only test it in the lab (the kernel version of Android devices typically lags 1-2 years behind), using a Linux device with kernels 5.4.0 and 5.5.6. We demonstrated consistent tracking, and moreover, as expected, the tracking was compatible with UDP source port -based tracking, i.e. the PRNG states obtained by the two techniques were associated.

Using IPID-based tracking, we also get consistent results with Mozilla Firefox, even on a 4-core device.

\end{document}